\documentclass[12pt,preprint]{aastex}

\newcommand{\be}{\begin{equation}}
\newcommand{\ee}{\end{equation}} 
\newcommand{\simless}{\lower.5ex\hbox{$\; \buildrel < \over \sim\;$}}
\newcommand{\simgreat}{\lower.5ex\hbox{$\; \buildrel > \over \sim\;$}}

\newcommand{\zhat}{{ {\hat z} }} 
 
\newcommand{\rhat}{{ {\hat r} }}

\begin{document}

\title{Effects of Turbulence on Cosmic Ray Propagation \\
in Protostars and Young Star/Disk Systems}

\author{Marco Fatuzzo} 
\affil{Physics Department, Xavier University, Cincinnati, OH 45207}
\email{fatuzzo@xavier.edu}
\and
\author{Fred C. Adams}
\affil{Physics Department, University of Michigan, Ann Arbor, MI 48109}
\email{fca@umich.edu}

\begin{abstract}

The magnetic fields associated with young stellar objects are expected
to have an hour-glass geometry, i.e., the magnetic field lines are
pinched as they thread the equatorial plane surrounding the forming
star but merge smoothly onto a background field at large distances.
With this field configuration, incoming cosmic rays experience both a
funneling effect that acts to enhance the flux impinging on the
circumstellar disk and a magnetic mirroring effect that acts to reduce
that flux. To leading order, these effects nearly cancel out for
simple underlying magnetic field structures. However, the environments
surrounding young stellar objects are expected to be highly turbulent.
This paper shows how the presence of magnetic field fluctuations
affects the process of magnetic mirroring, and thereby changes the
flux of cosmic rays striking circumstellar disks.  Turbulence has two
principle effects: 1) The (single) location of the magnetic mirror
point found in the absence of turbulence is replaced with a wide
distribution of values. 2) The median of the mirror point distribution
moves outward for sufficiently large fluctuation amplitudes (roughly
when $\delta B/B_0>0.2$ at the location of the turbulence-free mirror
point); the distribution becomes significantly non-gaussian in this
regime as well. These results may have significant consequences for
the ionization fraction of the disk, which in turn dictates the
efficiency with which disk material can accrete onto the central
object.  A similar reduction in cosmic ray flux can occur during the
earlier protostellar stages; the decrease in ionization can help
alleviate the magnetic braking problem that inhibits disk formation.

\end{abstract}


\section{Introduction}

Cosmic rays (CRs) significantly influence the physical properties of
the interstellar medium and are expected to play an important role in
the process of star formation. For example, the cosmic ray flux in
star forming regions directly affects the ionization levels
\citep{hayakawa,spitzer}, heating processes \citep{glassgold}, and
chemistry \citep{dalgarno} within the local environment.  On scales of
$\sim 0.1$ pc, ionization levels affect the coupling between the gas
and the magnetic fields, and in turn, the rate at which star formation
occurs (e.g., \citealt{fat06}; see also the reviews of
\citealt{sal87,mcos,crutcher}, and references therein).  On smaller
scales of $\sim 1-100$ AU, the ionization in circumstellar disks
impacts the extent of the disk where the magnetorotational instability
mechanism (MRI) remains active, and thereby mediates accretion rates
\citep{gammie}. Understanding how cosmic rays propagate through the
highly anisotropic and turbulent star forming environments thus
constitutes a fundamental problem in star formation theory.

It is well known that the motion of cosmic rays is strongly affected
by the structure of the magnetic field.  Specifically, large scale
field structures can both focus and mirror charged particles, and a
turbulent component that extends to scales smaller than the particle
gyration radius results in diffusive motion. All three effects are
expected to contribute in star formation environments.  Specifically,
the gravitational collapse of cores and subsequent formation of
protostellar disks are expected to produce hour-glass magnetic field
structures in which cosmic rays from the background environment can,
on the one hand, get funneled toward the central star/disk object,
and, on the other hand, eventually reflect away as they move into a
region of increasing magnetic field strength.  Note that the magnetic
field will attain an hour-glass geometry both in the limit of strong
fields, where the collapse is magnetically controlled, and for weak
fields, where the collapse flow drags in the field lines. In spite of
this ambiguity, observations indicate the presence of hour-glass-like
magnetic fields associated with protostars (see, e.g.,
\citealt{davidson}) and find alignment between the symmetry axis of the
(flattened) protostellar envelopes and the background magnetic fields
\citep{chapman}. In addition, young stellar objects are expected to be
highly dynamic and hence drive magnetic turbulence; the goal of this
paper is to ascertain how this turbulence affects the propagation of
cosmic rays into these systems.
 
Previous work has considered how cosmic rays propagate through the
magnetic field lines that thread molecular cores and related systems
\citep{skilling,cesarsky,chandran,padoan,desch,padogalli}, although
these analyses did not include the effects of turbulence. The results
of this previous work indicate that mirroring tends to dominate over
focusing, leading to a net reduction of the cosmic ray ionization rate
by a factor of $\sim 2$ -- 3 over most of a solar-mass core with
respect to the ``background" value for the intercloud medium (outside
the core).  Additional loss of cosmic ray flux can result from more
complicated field configurations due to twisting magnetic field lines
that are expected during protostellar collapse \citep{padohenngalli}.

In this paper, we explore the funneling and mirroring effects that
occur as cosmic rays move toward the circumstellar disks associated
with forming (or newly formed) stars, but also include the effects of
magnetic turbulence on cosmic ray propagation. We construct a new
nonstandard coordinate system to facilitate the analysis, which lends
itself more naturally to the underlying geometry (e.g., by allowing a
straightforward implementation of the required condition
$\nabla\cdot{\bf B}=0$).  As expected, we find that the presence of
turbulence leads to a distribution of possible outcomes for
essentially equivalent initial conditions. Specifically, there is no
longer a simple one-to-one correspondence between the initial
conditions of a comic ray and its mirroring point. In the presence of
turbulence, and for initial conditions where mirroring occurs far from
the disk, cosmic rays are equally likely to penetrate farther inward
or reflect earlier --- farther out --- compared to the turbulence-free
mirror radius. In other words, the distribution of mirroring points is
symmetric and centered on the value obtained without turbulence.
However, for conditions where the mirror points occur in the inner
regions near the disk, and for sufficiently large fluctuation 
amplitudes, turbulence acts primarily to enhance mirroring. As a
result, the net effect of turbulence is to increase the efficiency of
magnetic mirroring, i.e., turbulence acts to significantly reduce the
flux of cosmic rays that reach the circumstellar disk.

This paper is organized as follows. In Section 2, we construct a new
coordinate system, including the divergence operator, where one
coordinate follows the magnetic lines of the hour-glass-like
configuration. The perpendicular coordinates allow us to construct
magnetic field perturbations that point in the orthogonal directions
and are divergence-free. The specification of the magnetic field
perturbations is addressed in Section 3. Next we consider the
propagation of cosmic rays, in Section 4, including funneling and
mirroring in the absence of turbulence. Section 5 then includes the
effects of turbulence on cosmic ray propagation, and presents the
results from 120,000 numerical experiments. Finally, we conclude in
Section 6 with a summary of our results and a discussion of their
implications for star formation and disk accretion.

\section{Geometry} 

This section presents an idealized geometry for the magnetic field
extending from a young stellar object.  Near the star itself, we
expect the field to be dominated by a stellar dipole structure. But
the stellar wind will open up the field into a split monopole
configuration beyond some radius that is much larger than the stellar
radius, and much smaller than the radius of the circumstellar disk.
We therefore model the unperturbed (static) magnetic field extending
from the stellar system with a split-monopole component that
eventually merges with a uniform ``background'' field ${\bf B_\infty}
= B_\infty \hat z$.  This idealized form is expected to adequately
capture the important aspects of the underlying magnetic field
structure for the $z>0$ hemisphere beyond an inner boundary, which we
define through a radius $R$ (we specify the value below).  Note that incident cosmic rays that cross the
inner boundary are expected to have a high probability of interacting
with the circumstellar disk material.  The resulting hour-glass-like
geometry (for positive $z$) is then conveniently given by the
expression 
\be
{\bf B_0} = B_\infty \left[ {\varepsilon \over\xi^2} \rhat +  \zhat \right] \, ,
\label{field} 
\ee 
where $\xi = r/R$.  The value of $\varepsilon$ defines the relative
strength of the split-monopole component with respect to $B_\infty$,
and determines the approximate crossover radius $\xi_c =
\sqrt{\varepsilon}$ between the `nearly radial' and `nearly uniform'
regions of the magnetic field. 

With this configuration, the magnetic field is current-free and
curl-free, and can be written as the gradient of a scalar field. 
We define a scalar field $p$ that serves as the first field of
the coordinate system, i.e.,
\be
p = \xi \cos \theta - {\varepsilon\over \xi} \, ,
\label{pdef} 
\ee
where the gradient $\nabla p$ defines a vector field that points in the
direction of the magnetic field. We can then construct the
perpendicular vector field $\nabla q$ from a second scalar field
$q$ of the coordinate system, i.e., 
\be
q = {1 \over 2} \xi^2 \sin^2 \theta - \varepsilon \cos \theta \, . 
\label{qdef} 
\ee 
The pair $(p,q)$ thus represents a set of perpendicular coordinates in
the poloidal plane, with the azimuthal angle $\phi$ providing the
third scalar field of the coordinate system.  We note that the value
of $q$ remains constant on a given field line \citep{adams11,adgr12}.

The dimensionless covariant basis vectors $\underline{\epsilon}_j$ are 
given by the usual relations
\be
\underline{\epsilon}_p =  \,\nabla p, \qquad \underline{\epsilon}_q = \,\nabla q, 
\qquad {\rm and} \qquad \underline{\epsilon}_\phi = \,\nabla \phi\,,
\ee
where the gradient is written in terms of the variables ($\xi$, $\theta$, $\phi$).
Evaluating these quantities, we obtain
\be
\underline{\epsilon}_p =  
\left(\cos\theta+{\varepsilon\over \xi^2}\right)\hat r -\sin\theta\hat\theta
= {\varepsilon\over \xi^2}\hat r + \hat z = {{\bf B_0}\over B_\infty}\,,
\ee
\be
\underline{\epsilon}_q = 
\xi\sin\theta\left[\sin\theta\hat r 
+\left(\cos\theta+{\varepsilon\over\xi^2}
\right)\hat\theta\right]\,,
\ee
and
\be
\underline{\epsilon}_\phi = {1 \over \xi\sin\theta}\hat\phi\,.
\ee
We note that the quantities $\underline{\epsilon}_j$ are basis
vectors, rather than unit vectors, so that their length are not, in
general, equal to unity.  The corresponding unit vectors can trivially
be written as 
\be
\hat n_j = h_j \underline{\epsilon}_j\,,
\ee
where the corresponding scale factors 
$h_j = 1/|\underline{\epsilon}_j|$ are given by 
\be
h_p = \left[1+2\cos\theta{\varepsilon\over\xi^2}
+{\varepsilon^2\over\xi^4}\right]^{-1/2} = {B_\infty\over B_0}\,,
\ee
\be
h_q ={1\over\xi\sin\theta}\, \left[1+2\cos\theta{\varepsilon\over\xi^2}
+{\varepsilon^2\over\xi^4} \right]^{-1/2}\,,
\ee
and
\be
h_\phi = \xi\sin\theta\,.
\ee

In the limit of large $\xi$, the field lines point in the $\zhat$
direction.  The field lines that emanate radially outward from the
origin ($\xi = 0$) with angle $\theta_0$ thus map onto a cylinder at
large (spherical radii) $\xi$ with cylindrical radius $\varpi_\infty$. 
This radius is determined by the condition $q$ = {\sl constant}, i.e.,
\be
{1 \over 2} \xi^2 \sin^2 \theta - \varepsilon \cos \theta = 
- \varepsilon \cos \theta_0 \, , 
\ee
which can be rewritten in the form 
\be
\varpi_\infty^2  = 2 \varepsilon R^2 \left(1 - \cos \theta_0 \right) \, , 
\ee
where $\cos\theta_\infty \to 1$. The outermost radius $\varpi_{max}$
occurs for $\cos\theta_0$ = 0, i.e., the magnetic field line that
leaves from the equator of the central region, and the effective
feeding radius of the system is thus given by
\be
\varpi_{max} =  \left(2 \varepsilon \right)^{1/2} R \, . 
\label{feedrad} 
\ee

In order to relate the $(p,q,\phi)$ coordinate system with the more
traditional cartesian coordinate system, we show several field lines
in Figure 1 (solid curves) for the case that $\varepsilon = 10^4$.
``Equipotential'' lines of constant $p$ are also shown (dotted
curves).  We note that requiring $z \ge 0$ limits the coordinate $q$
to the range $-\varepsilon\le q \le 0$.  Furthermore, in the limit
that $\varepsilon \gg 1$, the inner boundary $\xi = 1$ is well
approximated by the $p = - \varepsilon$ line.

\begin{figure}
\figurenum{1}
{\centerline{\epsscale{0.90} \plotone{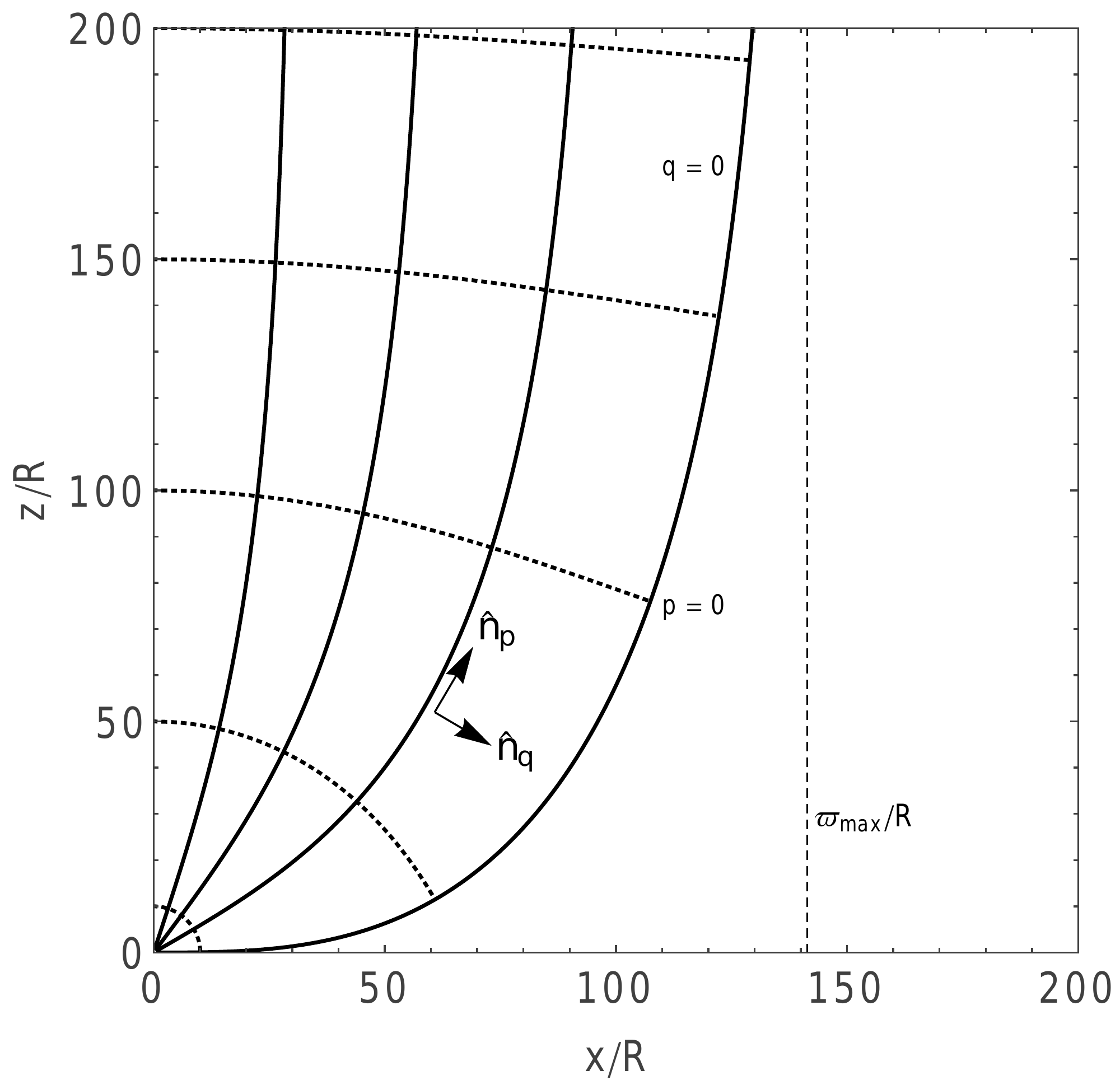} }}
\figcaption{Magnetic field geometry for $\varepsilon = 10^4$. 
Solid lines denote magnetic fields line, for which $q$ is constant.
The corresponding values of $q$ are, from the inner most line to the
outer most line, given by: $q = -9500, -8000, -5000$ and 0.  Dotted
lines denote ``equipotentials'' on which $p$ is constant.  Moving
outward, the coordinate has values  $p = -990$, --150, 0, 83 1/3, 
and 150.  The dashed line represents the cylindrical outer boundary at
$\varpi_{max}/R$.  Note that the unit vectors $\hat n_p$ and $\hat
n_q$ are not drawn to scale. }
\end{figure} 

\section{The Turbulent Magnetic Field}

At present, a complete theory of MHD turbulence in the interstellar
medium remains elusive. Nevertheless, it is generally understood that
turbulence is driven from a cascade of longer wavelengths to shorter
wavelengths as a result of wave-wave interactions. For strong MHD
turbulence in a uniform medium, this cascade seemingly produces eddies
on small spatial scales that are elongated in the direction of the
underlying magnetic field, so that the components of the wave vector
$k_\perp$ and $k_{||}$ are related by the expression $k_{||} \propto
k_\perp^{2/3}$ \citep{sg94,gs95,cho}. It is beyond the scope of this
paper to extend these results for our non-uniform geometry. Since our
aim here is to determine the possible effects of turbulence on cosmic
ray propagation into a star/disk system, we will assume a reasonable
form for the turbulent magnetic field as guided by basic principles.

Following the standard numerical approach for analyzing the
fundamental physics of ionic motion in a turbulent magnetic field, we
treat the total magnetic field ${\bf B}$ as a spatially turbulent
component $\delta{\bf B}$ superimposed onto the static hour-glass-like
background field ${\bf B_0}$ described in Section 2. The turbulent field
$\delta {\bf B}$ is generated by summing over a large number of
randomly polarized waves with effective wave vectors $\bar k_n$
logarithmically spaced between $\bar k_{min}$ and $\bar k_{max}$
(e.g., \citealt{giajok94,casse,sullivan,fat10}). We assume that each
term of the turbulent field is Alfv\'enic in the sense that $\delta
{\bf B_n} \perp {\bf B_0}$, and satisfies the no-monopole condition
$\nabla\cdot \delta {\bf B_n} = 0$.  Since the Alfv\'en speed $v_A$ is
much less than that of the relativistic cosmic rays, we can adopt a
static turbulent field for calculating the effects on particle motion.
This simplification then removes the necessity of specifying a
dispersion relation for each term.

Given these considerations, we assume a turbulent field of the form
\be
\delta {\bf B} = \sum_{n=1}^N A_n(p,q)  \,  \cos\left(\bar k_n\,p 
+ \beta_n\right) [ \cos\alpha_n \hat n_q + \sin\alpha_n
\hat n_\phi]\,,
\label{turbform} 
\ee
where the direction and phase of each term is set through a random
choice of $\alpha_n$ and $\beta_n$.  The values of $\bar k_1 = \bar
k_{min}$ and $\bar k_N = \bar k_{max}$ are defined in terms of a
maximum and minimum wavelength, as defined by the condition that $\bar
k_1 p$ and $\bar k_N p$ advance by $2 \pi$ as the distance along a
field line (as defined by its value of $q$) from the inner boundary
(i.e., $\xi = 1$) advances by $\lambda_{max}$ and $\lambda_{min}$,
respectively.  All other values of $k_n$ are then found through an
even logarithmic binning, with the total number of terms in the sum
given by $N = N_k \log_{10}[\lambda_{max}/\lambda_{min}]$, where $N_k$
is the number of waves desired per decade.  Following the results of 
previous studies (\citealt{fat10}; see also \citealt{evzweibel}), 
we set the number of waves per decade to $N_k = 25$. 

To illustrate how the resulting turbulent field will appear, we first
note that the values of $\bar k_1$ and $\bar k_N$ are, to a high level
of approximation, independent of the field line being considered when
$\lambda_{max} \ll \sqrt{\varepsilon} R$ and $\varepsilon \gg 1$.  The
first condition ensures that the field lines remain very nearly radial
as one follows a field line outward from then inner surface a distance
of one wavelength $\lambda_0$.  In so doing, the value of $\xi$
therefore changes from $1$ to $1+\lambda_0/R$, and the corresponding
change in $p$ is then given by 
\be
\Delta p = { \lambda_0\over R} \left[\cos\theta 
+{\varepsilon R\over \lambda_0+R}\right]
\approx { \varepsilon\lambda_0\over\lambda_0+ R}\,,
\ee
where the conditions $\lambda_0 \ll \sqrt{\varepsilon} R$ and
$\varepsilon \gg 1$ allow us to ignore the $\cos\theta$ term in the
final expression.  Since one wavelength corresponds to a change in the
argument $\bar k p$ of $2\pi$, one then finds 
\be
\bar k \approx {2\pi (\lambda_0+R)\over \varepsilon\lambda_0}\;.
\ee
To demonstrate how the wave profile changes along a field line, we
plot the function $g_q[p(\xi)] = \cos [\bar k\, p(\xi)]$ in Figure 2
for a wavelength of $\lambda_0 = 0.1 R$, where $p$ is evaluated as a
function of $\xi$ for a fixed value of $q$ (i.e., for a specified
field line).  We present results for the limiting values
$q=-\varepsilon$ and $q = 0$.  Clearly, the ``wave-like'' nature of
the turbulence, as defined by equation (\ref{turbform}), is the 
nearly same for all field lines near the inner surface. 

Likewise, the function $g_q[p(\xi)] \approx \cos[\bar k z/R]$ when
$\xi \gg \xi_c = \sqrt{\varepsilon}$, so that the wave-like nature of
the turbulence is the same for all field lines beyond the crossover
radius $\xi_c$. The wavelength $\lambda_\infty$ in this region is then
related to the wavelength at the inner boundary through the expression 
\be{
\lambda_\infty\over R}  = {\varepsilon \lambda_0\over \lambda_0 + R}\,.
\ee

\begin{figure}
\figurenum{2}
{\centerline{\epsscale{0.80} \plotone{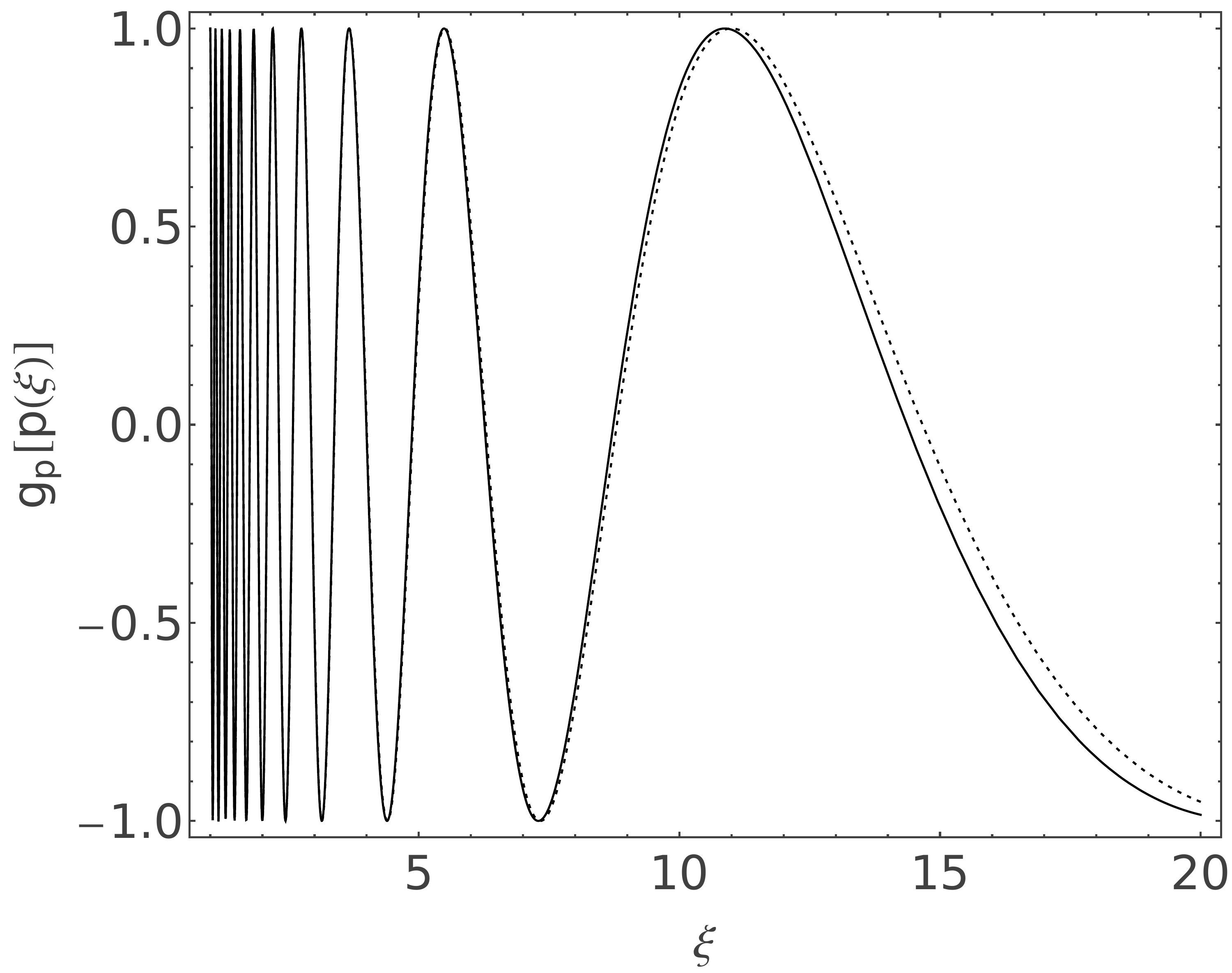} }}
\figcaption{The function $g_q[p(\xi)] = \cos [\bar k\, p(\xi)]$ for  
the case $\lambda = 0.1 R$, with the solid curve representing the case
$q = -\varepsilon$ and the dashed curve representing the case $q = 0$. } 
\end{figure}

In the cross-over region, we can characterize the wavelength $\lambda$
of the turbulence associated with the inner boundary wavelength
$\lambda_0$ through the condition that $\bar k p$ changes by $2\pi$ as
the radius changes from $\xi$ to $\xi + \lambda/R$ along the $q =
-\varepsilon$ field line (so that $p = \xi-\varepsilon/\xi$).  The
results are shown in Figure 3 for the values of $\lambda_0 = 0.1 R$
and $\lambda_0 = 10^{-5} R$.  As a point of reference, we also plot
the radius of gyration for a proton with Lorentz factor $\gamma =
10^2$ moving perpendicular to the background field ${\bf B_0}$, as
given by the expression 
\be
R_g = \gamma {m_p c^2 \over e B_0}\;,
\ee 
where we have assumed representative values of $B_\infty = 25\mu$G and
$R = 2\times 10^{12}$ cm (see discussion below).  Note that for such a
particle, the radius of gyration always falls within the range of
wavelengths spanning the turbulence profile generated by setting
$\lambda_{max} = 0.1 R$ and $\lambda_{min} = 10^{-5} R$.

\begin{figure}
\figurenum{3}
{\centerline{\epsscale{0.80} \plotone{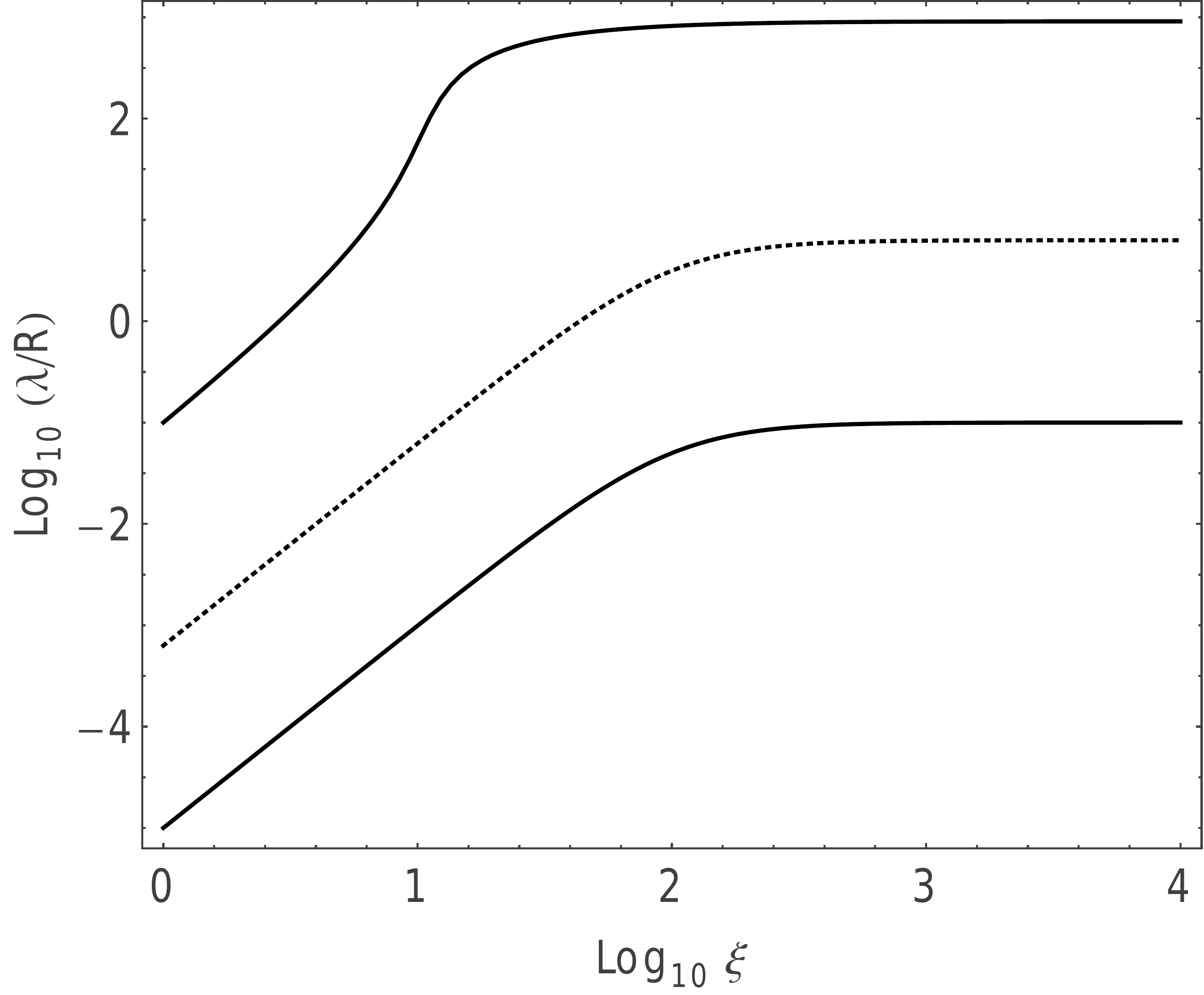} }}
\figcaption{The values of $\lambda$ as a function of $\xi$, as 
defined in the text, associated with $\lambda_0 = 0.1 R$ (upper solid
curve) and $\lambda_0=10^{-5}R$ (lower solid curve).  As a point of
reference, we also plot the value of $R_g/R$ as a function of $\xi$
(dotted curve), where $R_g$ is the radius of gyration for a proton
with Lorentz factor $\gamma=10^2$ moving perpendicular to the 
$q=-\varepsilon$ field line, for the assumed representative values of
$\varepsilon=10^4$, $B_\infty=25\mu$G, and $R=2\times 10^{12}$ cm. }
\end{figure}

To complete the analysis, we note that since the turbulent field is
axisymmetric, the divergence operator in our $(p,q,\phi)$ coordinate
system takes the form 
\begin{eqnarray}
\nabla\cdot \delta {\bf B_n} 
= {\cos(\bar k_n p+\beta_n) \cos(\alpha_n) \over h_p h_q h_\phi} 
\left[{\partial\over\partial q}  (h_p h_\phi A_n)\right]  = 0\,,
\end{eqnarray}
which requires
\be
h_p h_\phi A_n = f(p,\phi)\;.
\ee
Noting that $h_\phi = \xi \sin\theta$ and $h_p = B_\infty / B_0$, and
setting $f(p)$ equal to the constant $A_{0;n}$, one then finds 
\be
A_n(\xi,\theta) = {A_{0;n}\over \xi\sin\theta} {B_0\over B_\infty}\;,
\ee
where the desired spectrum of the turbulent magnetic field is set
through the appropriate choice of scaling for an assumed turbulent 
profile, i.e., 
\begin{equation}
A_{0;n}^2 = A_{0;1}^2\left[{k_n \over k_1} \right]^{-\Gamma}
{\Delta k_n\over \Delta k_1} 
= A_{0;1}^2\left[{k_n \over k_1} \right]^{-\Gamma+1}\,,
\end{equation}
where, e.g., $\Gamma = 3/2$ for Kraichnan and $\Gamma = 5/3$ for
Kolmogorov turbulence.  We note that for our logarithmic binning
scheme, the value of $\Delta k_n/k_n$ is the same for all values of
$n$.  The value of $A_{0;1}$ is set by an amplitude parameter $\eta$ 
that specifies the average energy density of the turbulent field with
respect to the background hour-glass field at the inner boundary; 
specifically, $\eta$ is defined through the expression  
\be
\eta = {\langle \delta B^2\rangle \over B_{0s}^2}\,,
\ee
where $B_{0s}$ is the magnitude of the hour-glass magnetic field at $\xi = 1$.

Figure 4 presents four different turbulent field lines produced using
four different values of $\eta$.  Moving outward, the amplitude
parameter $\eta=30$ for the $q=-9,500$ field line, $\eta=10$ for the
$q=-8000$ field line, $\eta=3$ for the $q=-5000$ field line, and
$\eta=1$ for the $q=0$ field line.  Note that for $\xi \ll \xi_c$, the
``nearly radial'' region of the background field, the magnitude of the
turbulent field scales as $\delta B \sim \sqrt{\eta} B_0 / \xi\sim
\sqrt{\eta}B_\infty \varepsilon / \xi^3$. On the other hand, for
$\xi\gg\xi_c$, the ``nearly uniform'' region of the background field,
the magnitude of the turbulent field scales as $\delta B\sim
\sqrt{\eta} B_\infty R/ \varpi_\infty$.

\begin{figure}
\figurenum{4}
{\centerline{\epsscale{0.80} \plotone{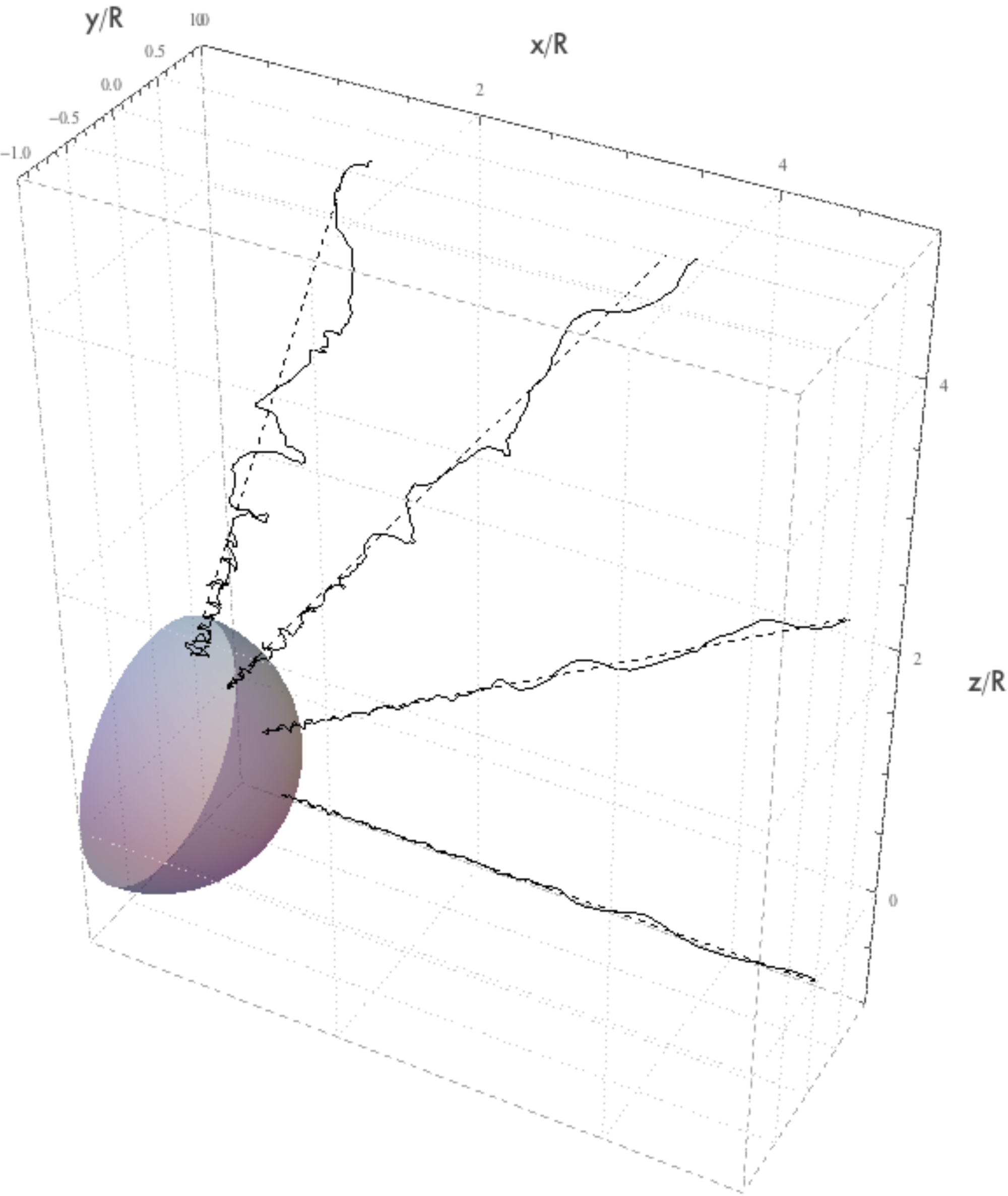} }}
\figcaption{Four different turbulent field lines produced using  
different values of the amplitude parameter $\eta$.  Moving outward,
$\eta=30$ for the $q=-9,500$ field line, $\eta=10$ for the $q=-8000$
field line, $\eta=3$ for the $q=-5000$ field line, and $\eta =1$ for
the $q=0$ field line. The gray surface denotes the inner boundary.  }
\end{figure}

\section{Basic Cosmic Ray Dynamics}

For clarity, we consider first the motion of relativistic charged
protons through the background hour-glass magnetic field without a
turbulent component \citep{padogalli}. The general equations that
govern the motion of protons with Lorentz factor $\gamma$ through a
magnetic field are 
\be
{d\over  dt}  (\gamma m_p {\bf v})=  
{e \,{\bf v} \times {\bf B}\over c} 
\qquad {\rm and} \qquad 
{d {\bf r}\over dt} = {\bf v}\;,
\ee
and are readily solved using standard numerical methods.

The magnetic moment of a relativistic proton is given by 
\be
\mu = {\gamma^2 m_p  v_\perp^2\over 2B}\;,
\ee
and is an adiabatic invariant under the condition that the field does
not change significantly within a cyclotron radius, i.e., in the limit 
\be
{\gamma m_p c v_\perp\over eB}  \ll {B\over |\nabla B|}\;,
\ee 
where $v_\perp$ is the component of the proton velocity perpendicular
to the magnetic field through which it is moving.  For the magnetic
field configuration used here, this limit is most stringent at the
inner surface, where it can be expressed in terms of a critical
Lorentz factor 
\be
\gamma\ll\gamma_{crit} \equiv 
{e\over m_p c^2}\left[{B^2\over |\nabla B|}\right]_R
={\varepsilon \over 2} {e B_\infty R\over m_pc^2}\,.
\ee

Since the Lorentz factor of the protons remains constant in a 
time-independent magnetic field, the adiabatic invariance can be
expressed as
\be
{\sin^2 \alpha_p\over B} = {\rm constant}\, ,
\ee 
where $\alpha_p$ is the pitch angle of a cosmic ray at a location
where the field strength is $B$.  As a cosmic ray moves toward the
inner radius, its pitch angle must increase to match the increasing
field strength; however, since $\sin\alpha_p \le 1$, the cosmic ray
must eventually reflect at a mirror point in the field.  For field
structures with $\varepsilon \gg 1$, cosmic rays initially far from
the cross-over region ($\xi \gg \xi_c$) must therefore have pitch
angles less than a maximum value if they are to penetrate the inner 
boundary at $\xi = 1$; this condition takes the form 
\be
\alpha_p < \alpha_{crit} = \sqrt{{B_\infty \over B_0(R)}} 
= {1\over\sqrt{ \varepsilon}}\, . 
\ee

In Figure 5, we plot the mirror point location (radius) as a function
of the cosine of the injection pitch angle for cosmic rays injected
into an hour-glass-like field with $\varepsilon = 10^4$. The figure
shows results for three different field lines (as defined by their
corresponding values of $q$), where all cases start from an initial
value of $z_i = 10^3 R$.

\begin{figure}
\figurenum{5}
{\centerline{\epsscale{0.80} \plotone{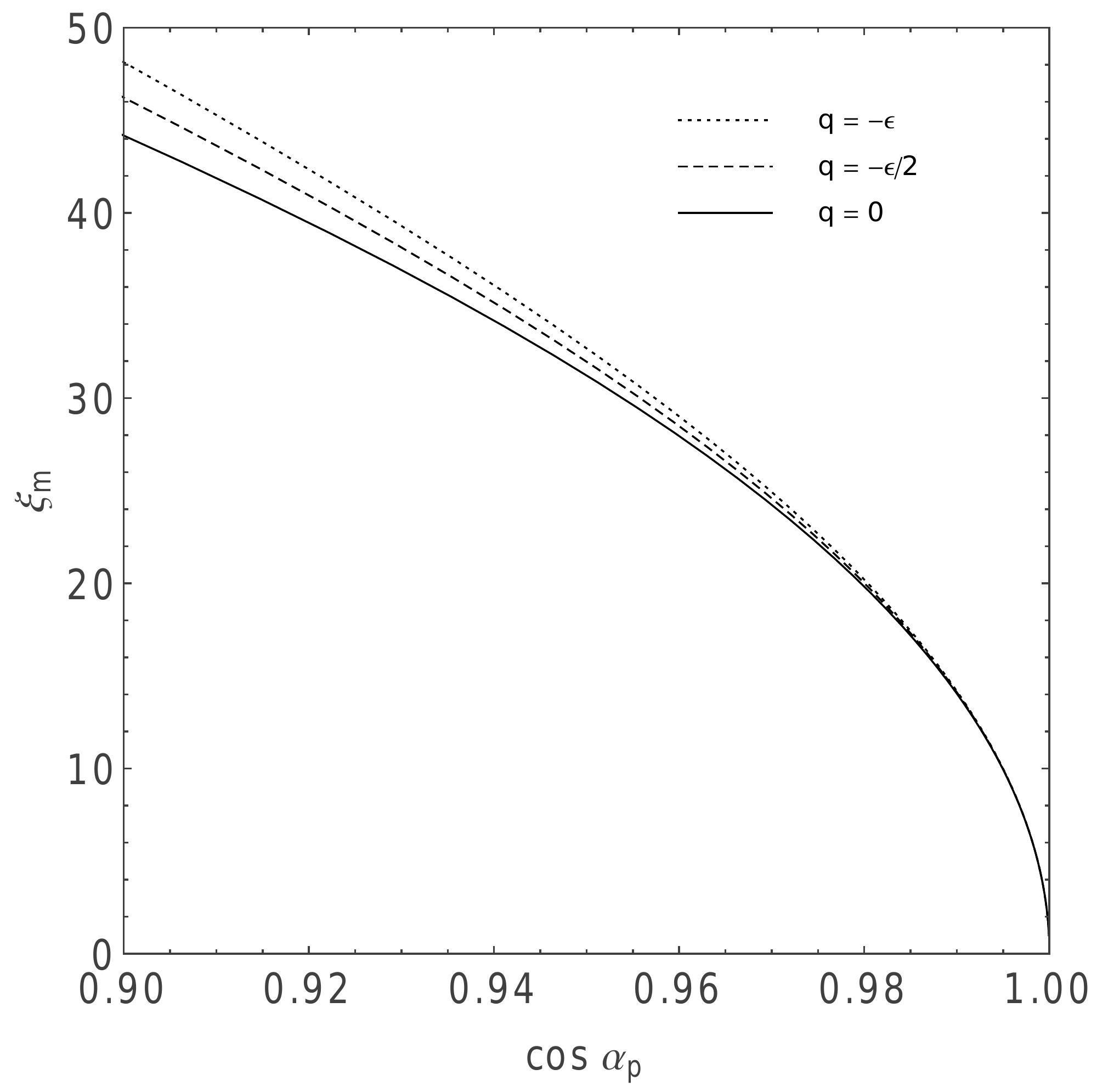} }}
\figcaption{The mirror radius $\xi_m$ as a function of the cosine 
of the injection inclination angle for comic-rays injected into an
hour-glass field with $\varepsilon = 10^4$ along three different field
lines corresponding to $q = 0$ (solid curve), $q = -\varepsilon/2$
(dashed curve) and $q = -\varepsilon$ (dotted curve); all trajectories 
start from an initial position with $z_i = 10^3 R$.  } 
\end{figure}

Figure 6 then shows the trajectories of three protons, each injected
at $z_i = 10^3 R$ with Lorentz factor $\gamma = 10^2$, into an
hour-glass field with $R = 2\times 10^{12}$ cm, $B_\infty = 25 \mu$G,
and $\varepsilon = 10^4$; the three cases correspond to the parameter
choices: 1) $\cos\alpha_p = 0.98$ and $q = -9500$, 2) $\cos\alpha_p =
0.99$ and $q = -5000$, and 3) $\cos\alpha_p = 0.995$ and $q = 0$.  The
mirror radii, as shown in this figure, are in excellent agreement with
the expected values as illustrated in Figure 5.  We note that for the
chosen field parameters, $\gamma \ll \gamma_{crit} = 8\times 10^4$.

\begin{figure}
\figurenum{6}
{\centerline{\epsscale{0.80} \plotone{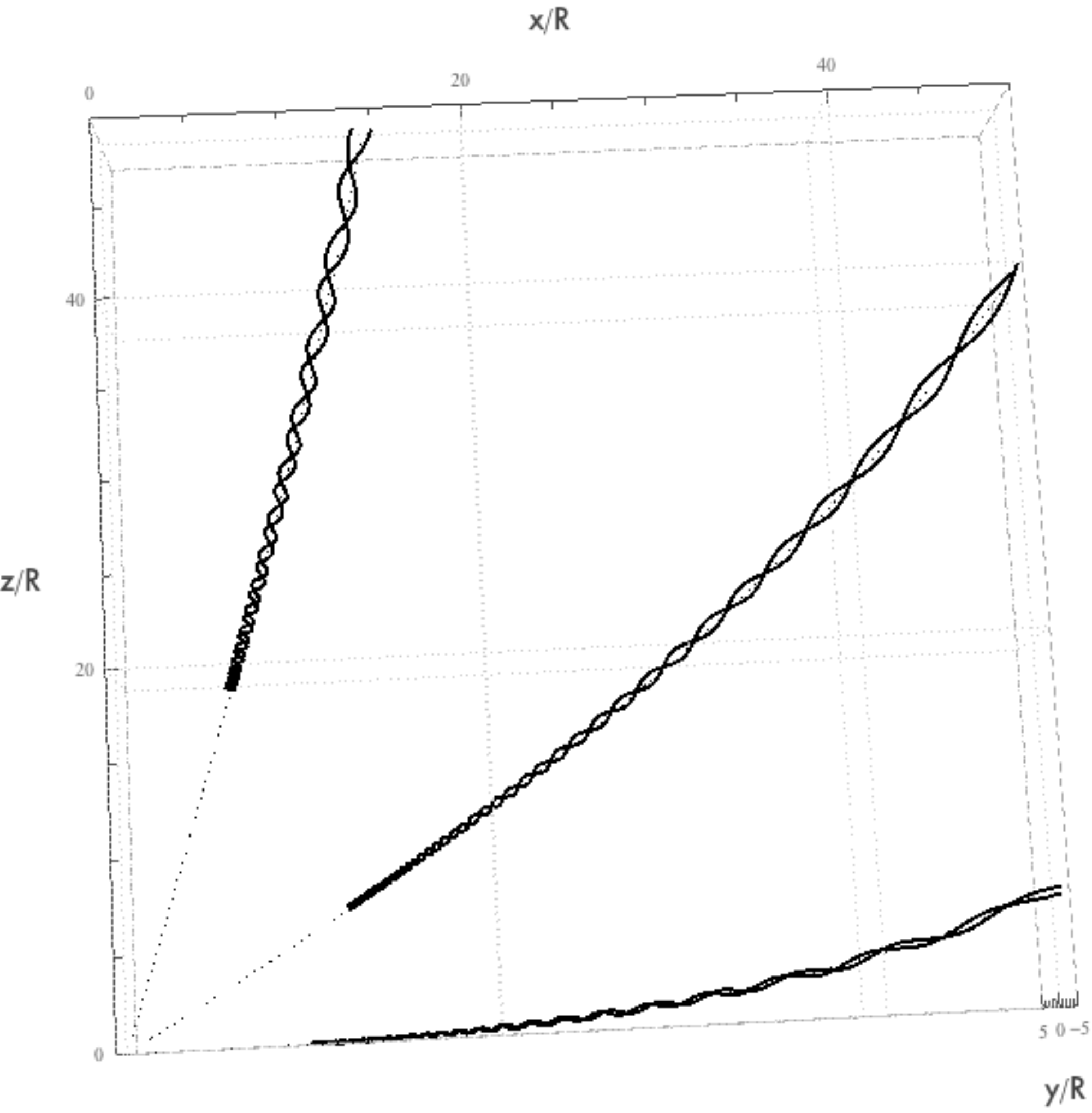} }}
\figcaption{Particle trajectories for three protons injected at 
$z_i = 10^4 R$ with Lorentz factors $\gamma=10^2$ into an hour-glass  
field with $R = 2\times 10^{12}$ cm, $B_\infty = 25 \mu$G, and
$\varepsilon = 10^4$.  Dashed lines show the magnetic field lines 
on which the particles were injected.  Left most trajectory: $q = -9500$,
$\cos\alpha_p = 0.98$.  Center trajectory: $q = -5000$, $\cos\alpha_p
= 0.99$.  Right most trajectory: $q = 0$, $\cos\alpha_p = 0.995$.  }
\end{figure}

If we assume that the velocity distribution of cosmic rays is
isotropic at distances much greater than $\xi_c$, then the fraction
${\cal F}$ of the cosmic ray flux that penetrates to the inner surface
($\xi = 1$) is given by 
\be
{\cal F}=1-\cos\alpha_{crit}=1-\left[1-\varepsilon^{-1}\right]^{1/2} 
\approx {1\over 2\varepsilon}\,.
\label{fraction} 
\ee
Clearly, only a small fraction of cosmic rays penetrate all the way to
the depth $\xi = 1$, with the remainder being mirrored back. However,
this apparent reduction in the cosmic ray flux impinging upon the
stellar disk is offset by the funnel effect resulting from the
hour-glass geometry.  In the limit of large $\xi$, the field lines
point in the $\zhat$ direction; as a result, field lines that cross
the inner boundary (at $\xi = 1$) can be mapped onto a cylinder at
large $\xi$.  Using the effective feeding radius of the system, as
given by equation (\ref{feedrad}), the effective input area 
$A_{\rm eff}$ of the system (the area from which cosmic rays are
harvested from the background medium) is thus given by 
\be
A_{\rm eff} = 2 \pi \varepsilon R^2\, . 
\ee
The cosmic rays are thus funneled from an initial area $A_{\rm eff}$
to an inner region with cross-sectional area $A_s = \pi R^2$, which
enhances the cosmic ray flux by a factor ${\cal E}=2\varepsilon$.  
The net factor by which the cosmic ray flux changes is thus given by 
\be
f = {\cal F}{\cal E}= 1 \, . 
\ee
In other words, to leading order, the mirror effect and the funnel
effect cancel out (in agreement with previous treatments, e.g.,
\citealt{padogalli}). We note that a simple flux-freezing argument
gives a similar cancellation between the mirror effect and the funnel
effect.

\section{Effects of Turbulence}

This section generalizes the calculation of the previous section to
include turbulent fluctuations of the magnetic field, and shows how
cosmic ray propagation can be affected. In the presence of turbulence,
charged particles interact resonantly with the magnetic field, and are
most strongly influenced by field fluctuations with wavelength
$\lambda\sim R_g$.  As a result, the magnetic moment of a cosmic ray
is no longer invariant.  This point is illustrated in Figure 7, which
plots the product $(B_\infty/B)\sin^2\alpha$ as a function of $\xi$
for a proton injected toward the inner boundary on the $q=-5000$ field
line with a pitch angle defined by $\cos\alpha_p=0.99$; results are
shown for four different turbulence levels, with amplitude parameter
$\eta$ = 0, 0.1, 1, and 10. As expected, the presence of magnetic
turbulence can displace the location at which mirroring occurs.

\begin{figure}
\figurenum{7}
{\centerline{\epsscale{0.80} \plotone{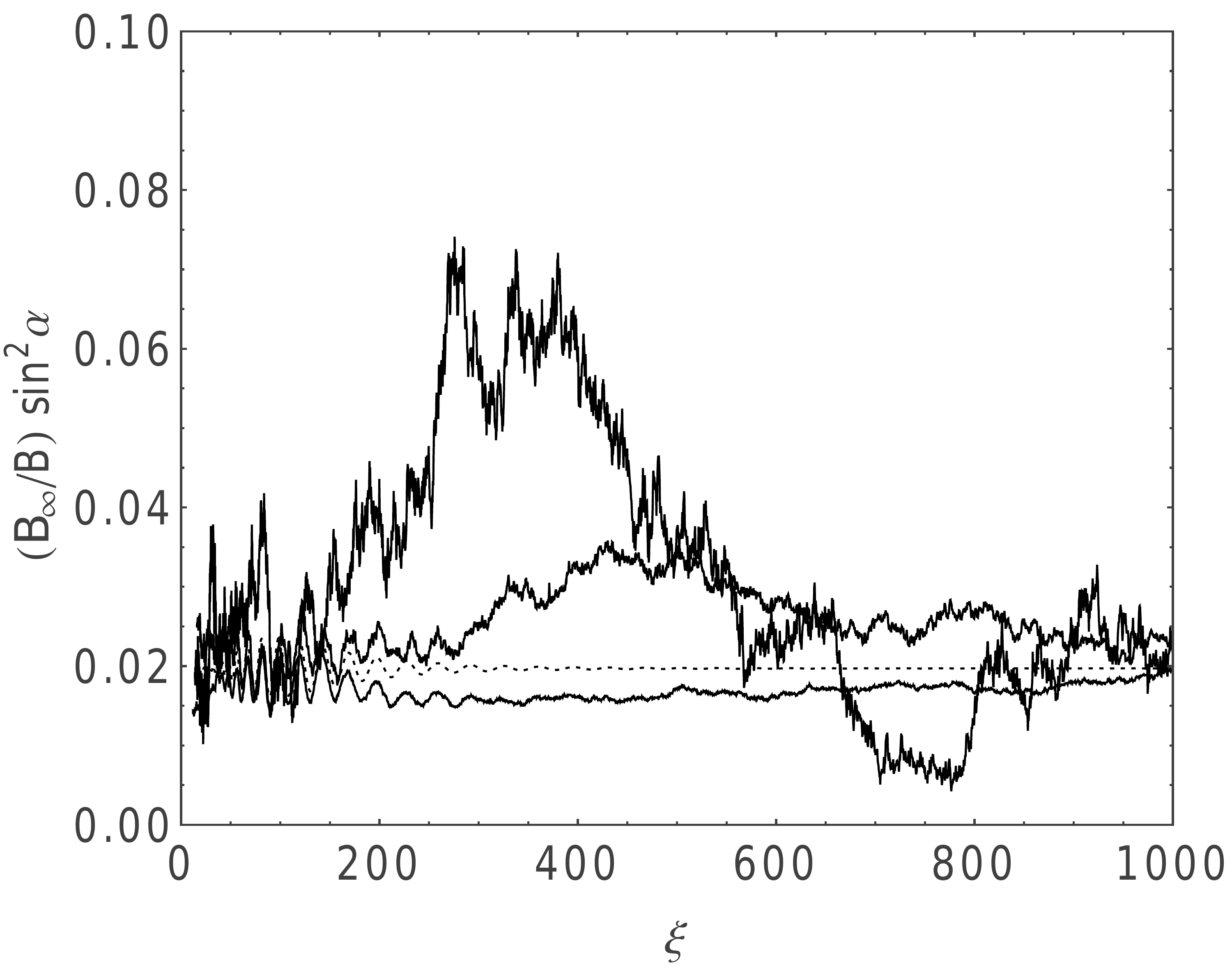} }}
\figcaption{The value of $(B_\infty/B)\sin^2\alpha$ as a function of
$\xi$ for a protons injected on the $q = -5000$ field line toward the  
inner boundary with an initial pitch angle corresponding to
$\cos\alpha_p$ = 0.99. Results are shown for four different turbulence 
levels.  The dotted line shows the result for amplitude parameter
$\eta = 0$ (no turbulence).  The other curves represent, in order of
increasing variability, $\eta$ = 0.1, 1, and 10. }
\end{figure}

Although the governing equations are deterministic, the motion of
charged particles through a turbulent magnetic field is chaotic in
nature. As a result, a complete analysis requires a statistical
approach. Toward that end, we have carried out a large ensemble of
numerical experiments to study cosmic ray propagation through a
turbulent magnetic field.  We define a single experiment as a
numerical investigation of the particle dynamics through a given type
of turbulent environment, starting with a given particle injection
scenario, as described below. 

For all experiments, we adopt fiducial values of $B_\infty = 25\mu$G,
$R = 2\times 10^{12}$ cm, and $\varepsilon = 10^4$ for the background
magnetic field, and we use
$\lambda_{max} = 0.1 R$, and $\lambda_{min} = 10^{-5} R$ for the
turbulent field.  We focus primarily on Kolmogorov turbulence ($\Gamma = 5/3$),
but perform an experiment using Kraichnan turbulence ($\Gamma = 3/2$)
for comparison.  The turbulence level, as defined by $\eta$, is one
of the experimental parameters.  Particles are injected toward the
origin from $z_i = 10^3 R$ with one of the following injection
scenarios: 1) all particles are injected with the same inclination
angle $\alpha_p$ and from the same field line, as specified by $q$; 2)
all particles are injected with the same inclination angle $\alpha_p$,
but randomly distributed throughout the portion of the $x-y$ plane (at
fixed height $z = z_i$) that funnels particles directly to the inner
surface; 3) same as scenario 2, but with particles injected with
inclination angles drawn randomly from a flat (uniform) $\mu_p =
\cos\alpha_p$ distribution between $0.9999$ and 1.  For each
experiment, we numerically integrate the equations of motion for
$N_p=10^4$ monoenergetic protons (as defined by their Lorentz factor
$\gamma$) until reflection occurs, with each particle sampling its own
unique realization of the magnetic turbulence through a random
selection of the values for $\alpha_n$ and $\beta_n$.  The radius
$\xi_m$ where mirroring occurs is the output of each particle run, and
the median value $\bar\xi_m$ and normalized width $\sigma_\xi$ of the
ensuing distributions then serve as the output measures for a given
experiment.  The experiments and corresponding output measures are
summarized in Table 1, and the distributions of mirroring radii for
each experiment are shown in Figures 8 -- 19.  Note that the values of
$\bar\xi_m$ are given in terms of the median for the corresponding
distribution one would obtain in the absence of turbulence, which is
denoted as $\bar\xi_{m0}$, whereas the values of $\sigma_\xi$ are
given in terms of $\bar\xi_m$.

The general effect that turbulence has on mirroring is illustrated by
the results of Experiments 1 -- 3, as shown in Figures 8 -- 10.  For
these experiments, all particles were injected from the same location
and with the same pitch angle, but the magnetic environments had
different turbulence levels.  Not surprisingly, the distributions
broaden as the turbulence strength parameter $\eta$ increases.
Interestingly, the distribution of mirroring radii for the $\eta =
0.1$ and $\eta = 1$ cases are well represented by normal distributions
whose median values (dashed lines) are very nearly equal to the
mirroring radius in the absence of turbulence (solid line).  For the
$\eta = 10$ case, the distribution starts to deviate from normal and
has a median value that is significantly greater than the
turbulence-free mirroring radius.  Note that for this latter case, the
ratio of the turbulent to underlying field magnitudes at the location
of the turbulent free mirror point scales as $\delta B / B_0 \sim
\sqrt{\eta}/\bar \xi_{m0} \sim 0.2$, suggesting that ``strong''
turbulence enhances the mirroring effect, thereby reducing the ability
of charged particles to penetrate into regions of increasing magnetic
fields.  In contrast, the corresponding ratios for the $\eta = 0.1$
and $\eta = 1.0$ cases are $\delta B / B_0 \sim 0.02$ and $\delta B /
B_0 \sim 0.07$, respectively.

The results of Experiments 2, 4 and 5 illustrate how reflection is
affected by the field line on which a particle moves.  As shown by
Figures 9, 11 and 12, there is little difference between the
distributions of mirroring radii for particles spiraling inward along
the $q=-9500$, $q=-5000$, and $q=0$ field lines.  We note, also, that
the turbulence-free mirror points along different field lines converge
as $\xi\rightarrow 1$, as illustrated in Figure 5.  As a result, there
is no need to weight the distributions of initial positions when
considering the overall effects of funneling and mirroring; this
finding validates our use of random starting positions along the $x-y$
plane for injection scenarios 2 and 3, as defined above.

How turbulence affects mirroring for particles able to penetrate
further into the field as a result of a smaller injection pitch angle
is illustrated by the results of Experiments 6 -- 8, each of which
adopts injection scenario 2.  Consistent with the results of
Experiments 1 -- 3, turbulence is seen to enhance mirroring in the
``strong'' turbulence limit.  Indeed, the ratios of the turbulent to
underlying field magnitudes at the location of the turbulent-free
mirror point for Experiments 6 -- 8 are given by $\delta B/B_0\sim
0.07$, $\delta B / B_0 \sim 0.02$ and $\delta B / B_0 \sim 0.7$
respectively.  We also note that the distributions obtained for both
Experiments 7 and 8 are quite similar, and are both characterized by
median values of $\bar\xi_m \approx 8$.  This result suggests that
once a certain turbulence threshold $\delta B / B_0\sim 1/8$ is
reached, particles are effectively mirrored.

Experiments 9 -- 13 explore how effective turbulence is likely to be
at limiting the number of cosmic rays reaching the star/disk
system.  Particle injection scheme 3 is adopted, for which half the
particles are injected with a small enough pitch angle to reach the
inner boundary in the absence of turbulence, as indicated by the
distribution shown in Figure 16.  As is clearly seen from Figures 17
-- 20, the presence of turbulence significantly reduces the number of
cosmic rays that reach the star/disk system, though cosmic rays with a
smaller energy do seem to be more likely to do so.  In addition,
Kraichnan turbulence ($\Gamma = 3/2$) appears to be slightly more effective than
Kolmogorov tubulence ($\Gamma = 5/3)$ at
limiting the number of cosmic rays that reach the star/disk system.
This result is consistent with the fact that Kraichnan turbulence has more
power at shorter wavelengths, and therefore can more effectively scatter
lower energy particles.
Finally, we note, 
that the ratios of the turbulent to underlying field magnitudes were
mirroring occurs (as characterized by the median of the distribution),
is $\delta B / B_0 \sim 0.1$, $\delta B / B_0 \sim 0.1$,  $\delta B
/ B_0 \sim 0.2$, and $\delta B
/ B_0 \sim 0.1$ for Experiments 10, 11, 12 and 13, respectively.

\begin{deluxetable}{rcccccccc}
\tablecolumns{9}
\tablewidth{0pc}
\tablecaption{\bf Summary of Experiments}
\tablehead{
\colhead{Exp} & \colhead{IS}   & \colhead{$\eta$}    &\colhead{$\Gamma$}
& \colhead{$\gamma$} & \colhead{$\cos\alpha_p$}  &  \colhead{$q$}&
\colhead{$\bar\xi_m/\bar\xi_{m0}$}& \colhead{$\sigma_\xi / \bar\xi_m$}}
\startdata
1 & 1  & 0.1 & 5/3 & $10^2$ &0.99 & --5000 &1.0& 0.08 \\
2 & 1  & 1 & 5/3 & $10^2$ &0.99 & --5000 &1.0 &0.26 \\
3 & 1  & 10 & 5/3 & $10^2$ &0.99 & --5000 &1.2& 0.41\\
4 & 1  & 1 & 5/3 & $10^2$ &0.99 & --9500 &1.0&0.29 \\
5 & 1  & 1 & 5/3 & $10^2$ &0.99 & 0 &1.0&0.23\\
6 & 2  & 1 & 5/3 & $10^2$ &0.99 & $-$ &1.0 &0.28\\
7 & 2  & 1 & 5/3 & $10^2$ &0.999 & $-$ &1.8&0.41\\
8 & 2  & 1 & 5/3 & $10^2$ &0.9999 & $-$ &5.8&0.36\\
9 & 3  & 0 & 5/3 & $10^2$ &$-$ & $-$ &1.0& $-$\\
10 & 3  & 0.1 & 5/3 & $10^2$ &$-$ & $-$ &3.1& $-$\\
11 & 3  & 1 & 5/3 & $10^2$ &$-$ & $-$ &8.2& $-$\\
12 & 3  & 1 & 5/3 & $10$ &$-$ & $-$ &5.8& $-$\\
13 & 3  & 1 & 3/2 & $10^2$ &$-$ & $-$ &9.1& $-$\\
\enddata
$\,$ \hfil 
\noindent
$\,$  
Table 1: The columns give the values of the experiment number (Exp),
the scheme for initial conditions (IS), 
the amplitude parameter for fluctuations ($\eta$),  the turbulence profile
parameter ($\Gamma$),
the Lorentz factor ($\gamma$), the starting
injection angle ($\alpha_p$), the coordinate that labels the field
line ($q$), the ratio of the median mirroring point to that obtained
with no turbulence ($\bar\xi_m/\bar\xi_{m0}$), and finally the
normalized width of the distribution of mirror point radii
($\sigma_\xi / \bar\xi_m$). 
\end{deluxetable}

\begin{figure}
\figurenum{8}
{\centerline{\epsscale{0.80} \plotone{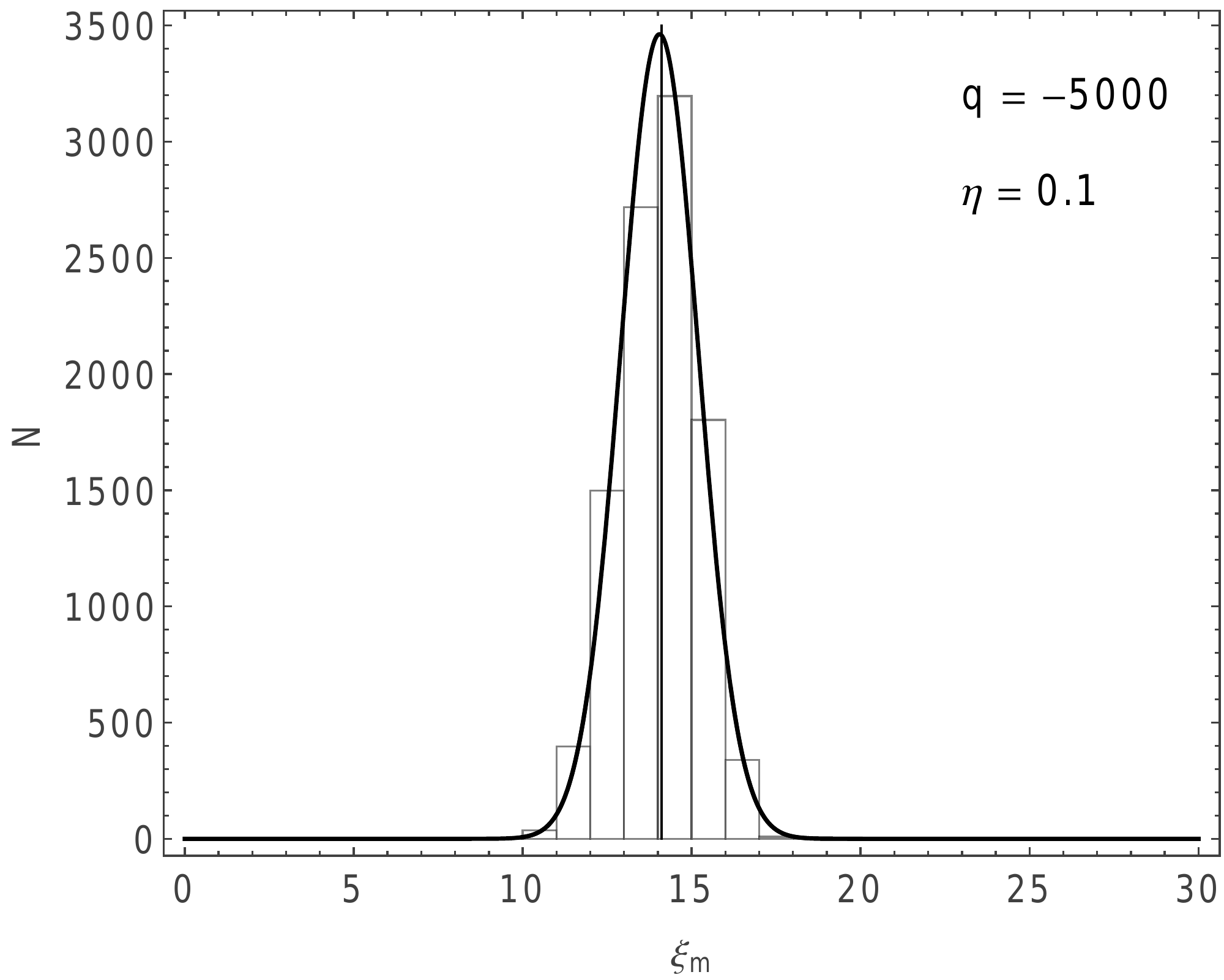} }}
\figcaption{Result of Experiment 1, as presented through the distribution 
of mirror radii for $10^3$ particles injected along the $q = -5000$
field line, starting with $z_i = 10^3 R$, Lorentz factor $\gamma =
10^2$, and with pitch angle $\cos \alpha_p = 0.99$.  The background
field $B_0$ is defined through the parameter $\varepsilon = 10^4$, and
the turbulent magnetic field strength is set at $\eta = 0.1$.  The
solid line denotes the value $\xi_{m0}$ of the mirror radius in
absence of turbulence, and the dashed line denotes the median value
$\bar\xi_m$ of distribution. }
\end{figure}

\begin{figure}
\figurenum{9}
{\centerline{\epsscale{0.80} \plotone{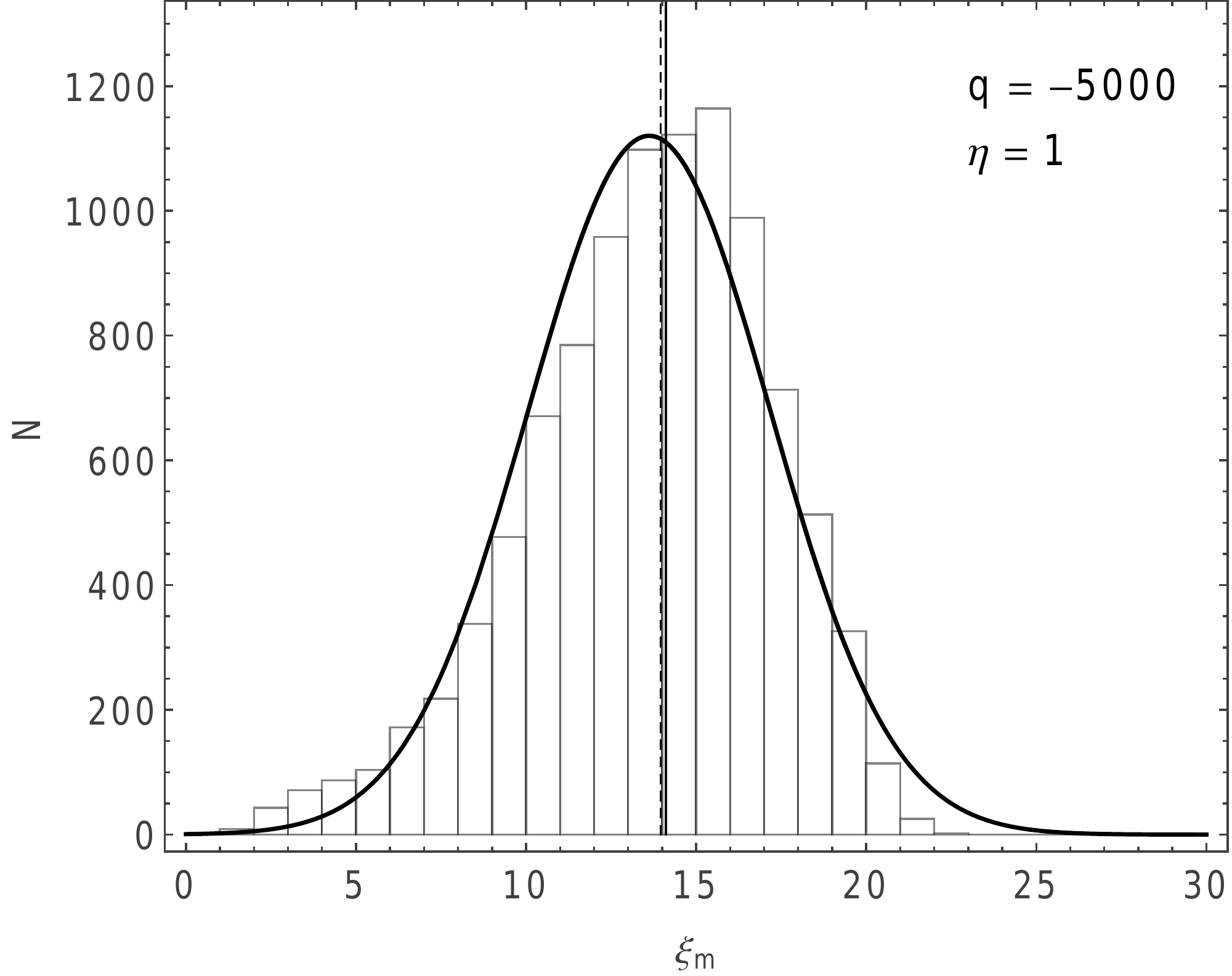} }}
\figcaption{Same as Figure 8, but for Experiment 2, for which $\eta = 1$.} 
\end{figure}

\begin{figure}
\figurenum{10}
{\centerline{\epsscale{0.80} \plotone{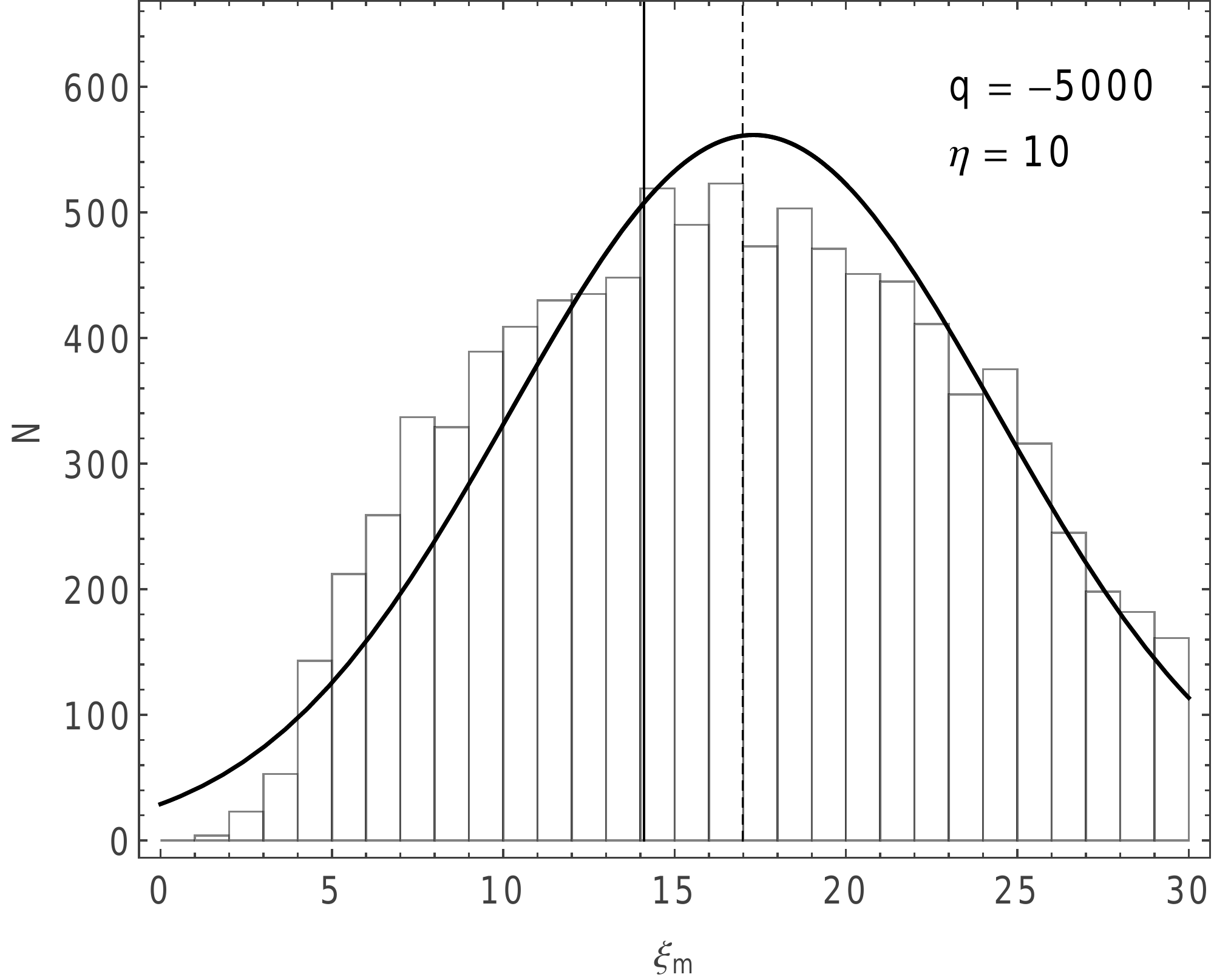} }}
\figcaption{Same as Figure 8, but for Experiment 3, for which $\eta = 10$.} 
\end{figure}

\begin{figure}
\figurenum{11}
{\centerline{\epsscale{0.80} \plotone{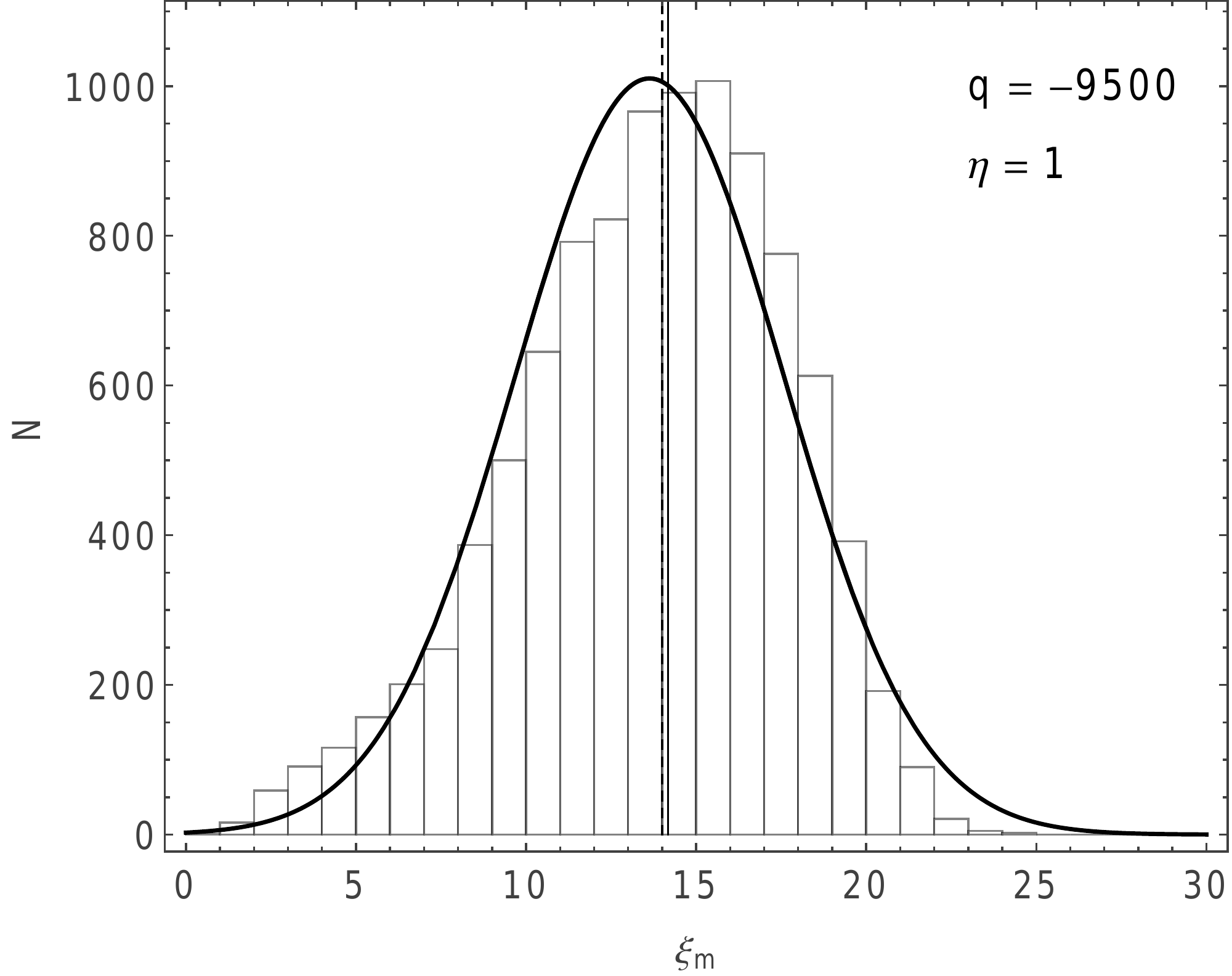} }}
\figcaption{Same as Figure 8, but for Experiment 4, for which $\eta = 1$ and $q = -9500$.} 
\end{figure}

\begin{figure}
\figurenum{12}
{\centerline{\epsscale{0.80} \plotone{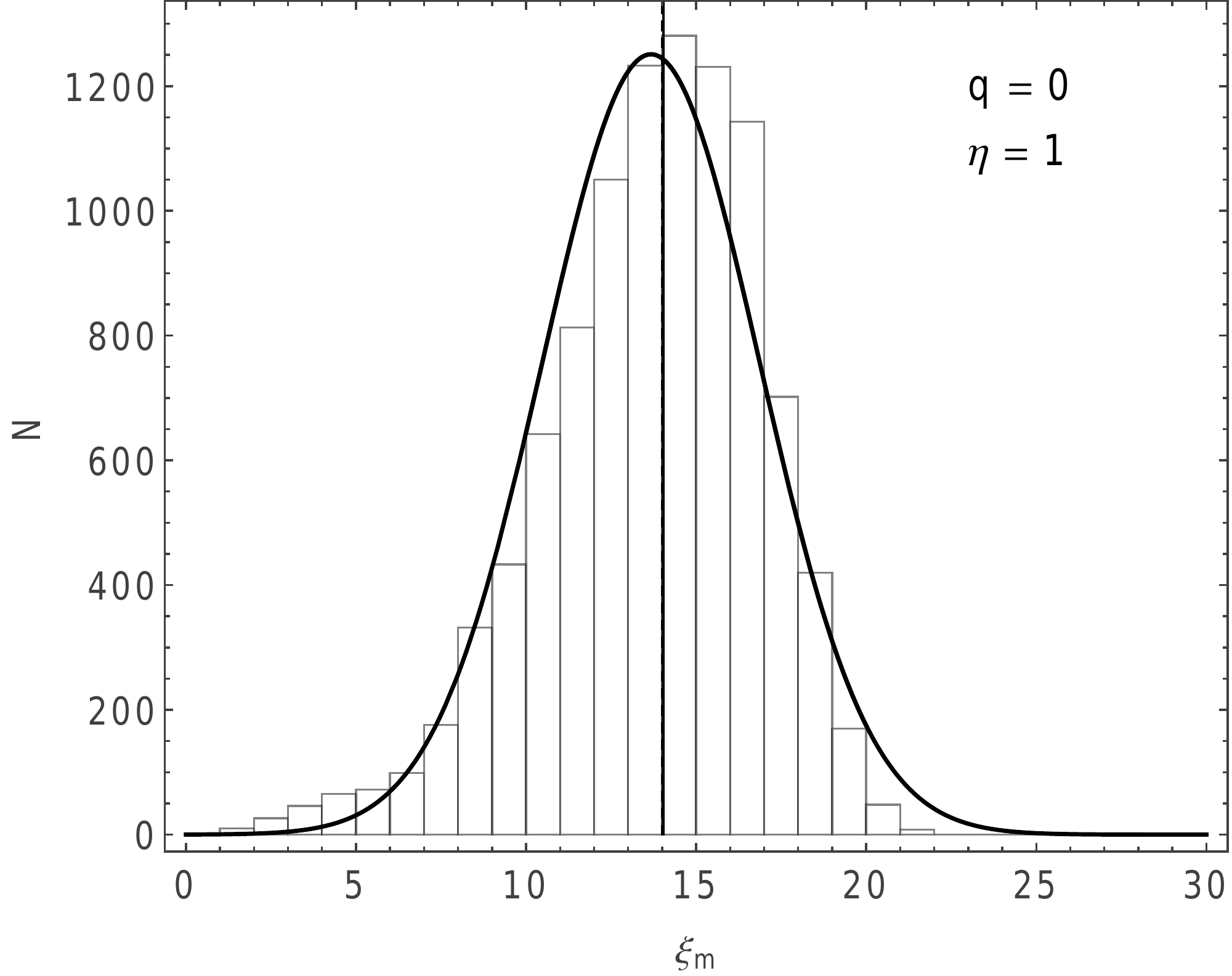} }}
\figcaption{Same as Figure 8, but for Experiment 5, for which $\eta = 1$ and $q = 0$.} 
\end{figure}

\begin{figure}
\figurenum{13}
{\centerline{\epsscale{0.80} \plotone{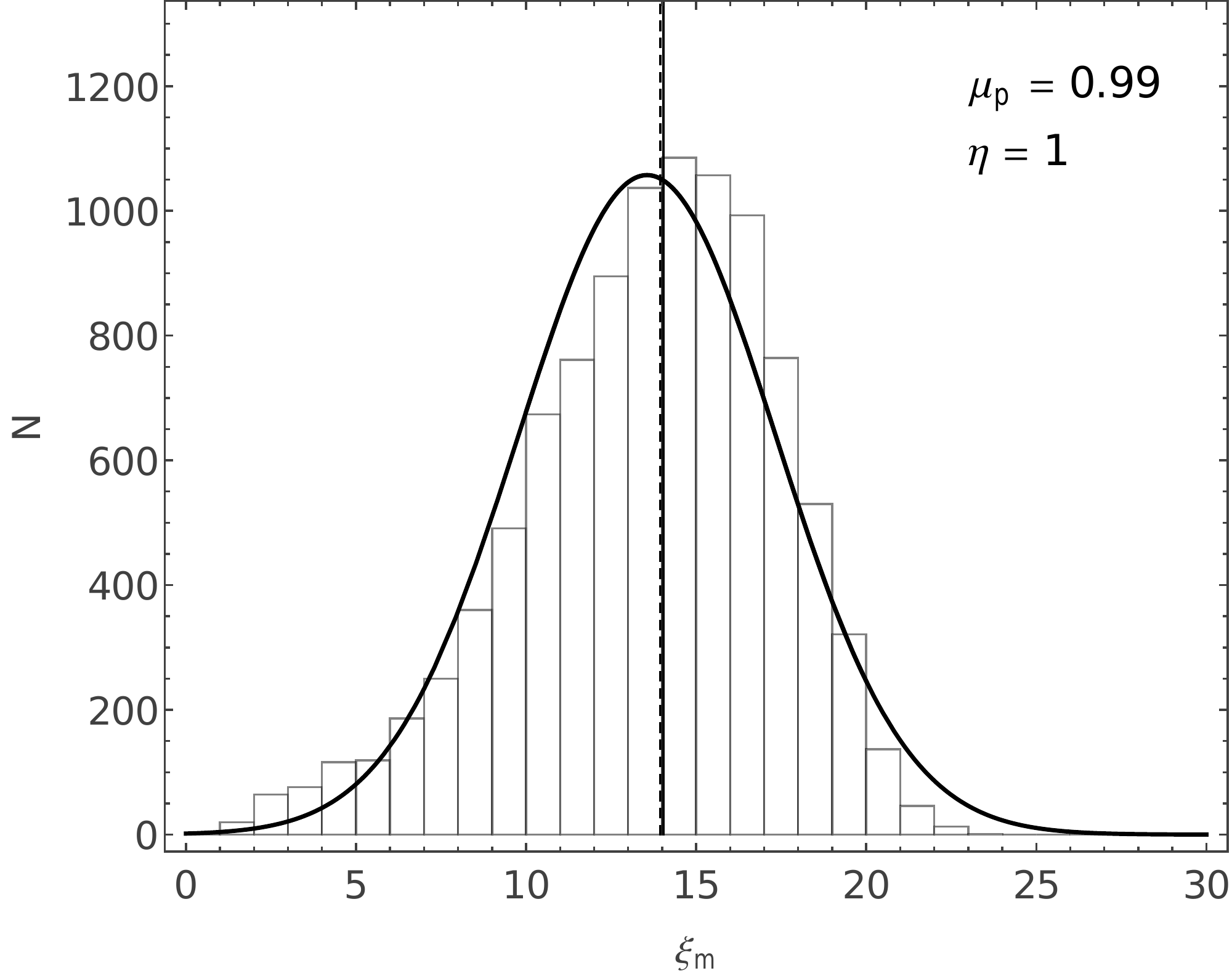} }}
\figcaption{Result of Experiment 6, as presented through the  
distribution of mirror radii for $10^3$ particles with Lorentz factor
$\gamma = 10^2$ and injected with pitch angle $\cos \alpha_p = 0.99$
from random locations at $z_i = 10^3 R$ from the part of $x-y$ plane
that funnels particles directly onto the inner surface.  The
background field $B_0$ is defined through the parameter $\varepsilon =
10^4$, and the turbulent magnetic field strength is set at $\eta = 1$.
The solid line denotes the value $\xi_{m0}$ of the mirror radius in
absence of turbulence, and the dashed line denotes the median value
$\bar\xi_m$ of distribution.  }
\end{figure}

\begin{figure}
\figurenum{14}
{\centerline{\epsscale{0.80} \plotone{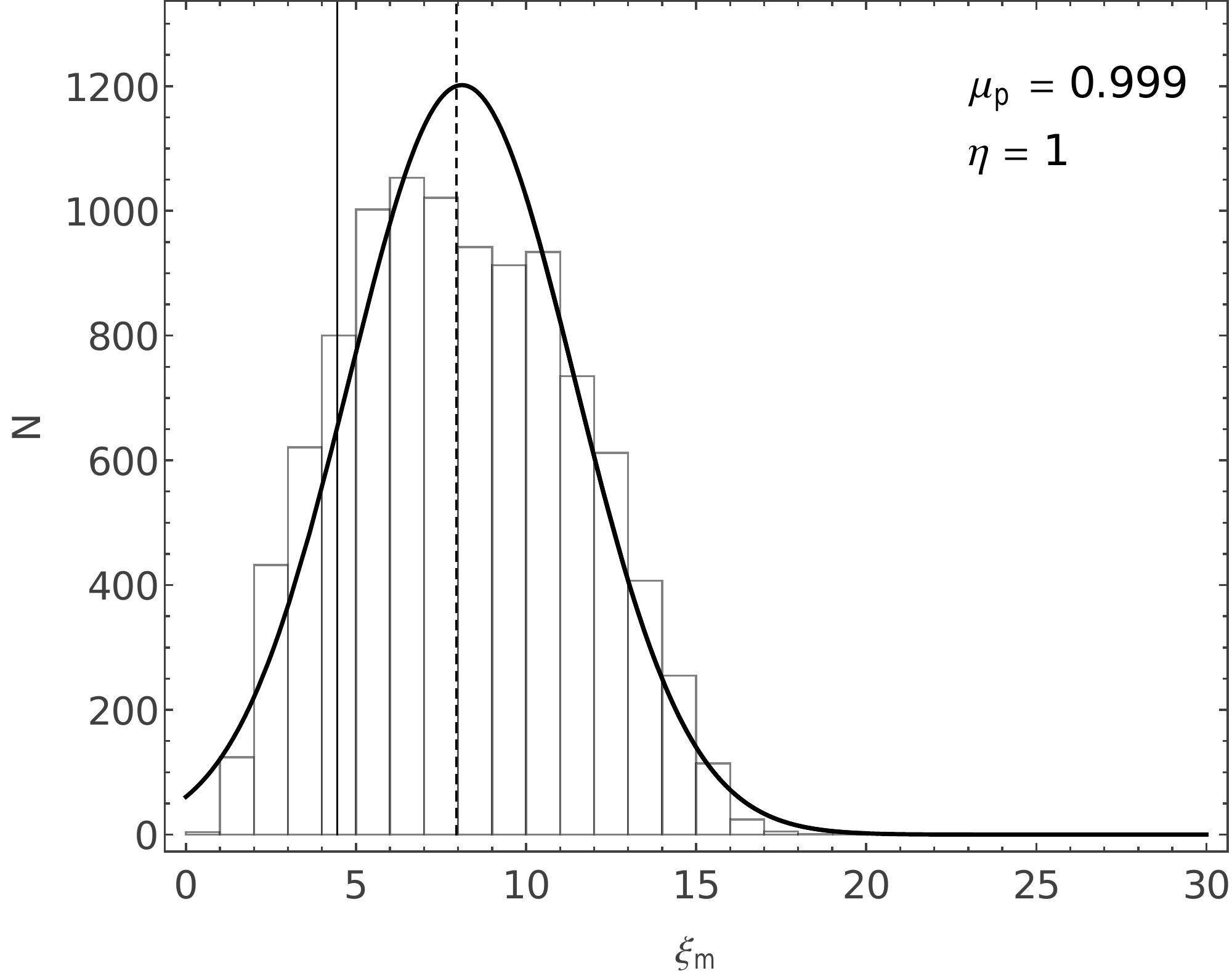} }}
\figcaption{Same as Figure 13, but for Experiment 7, for which $\cos\alpha_p = 0.999$.} 
\end{figure}

\begin{figure}
\figurenum{15}
{\centerline{\epsscale{0.80} \plotone{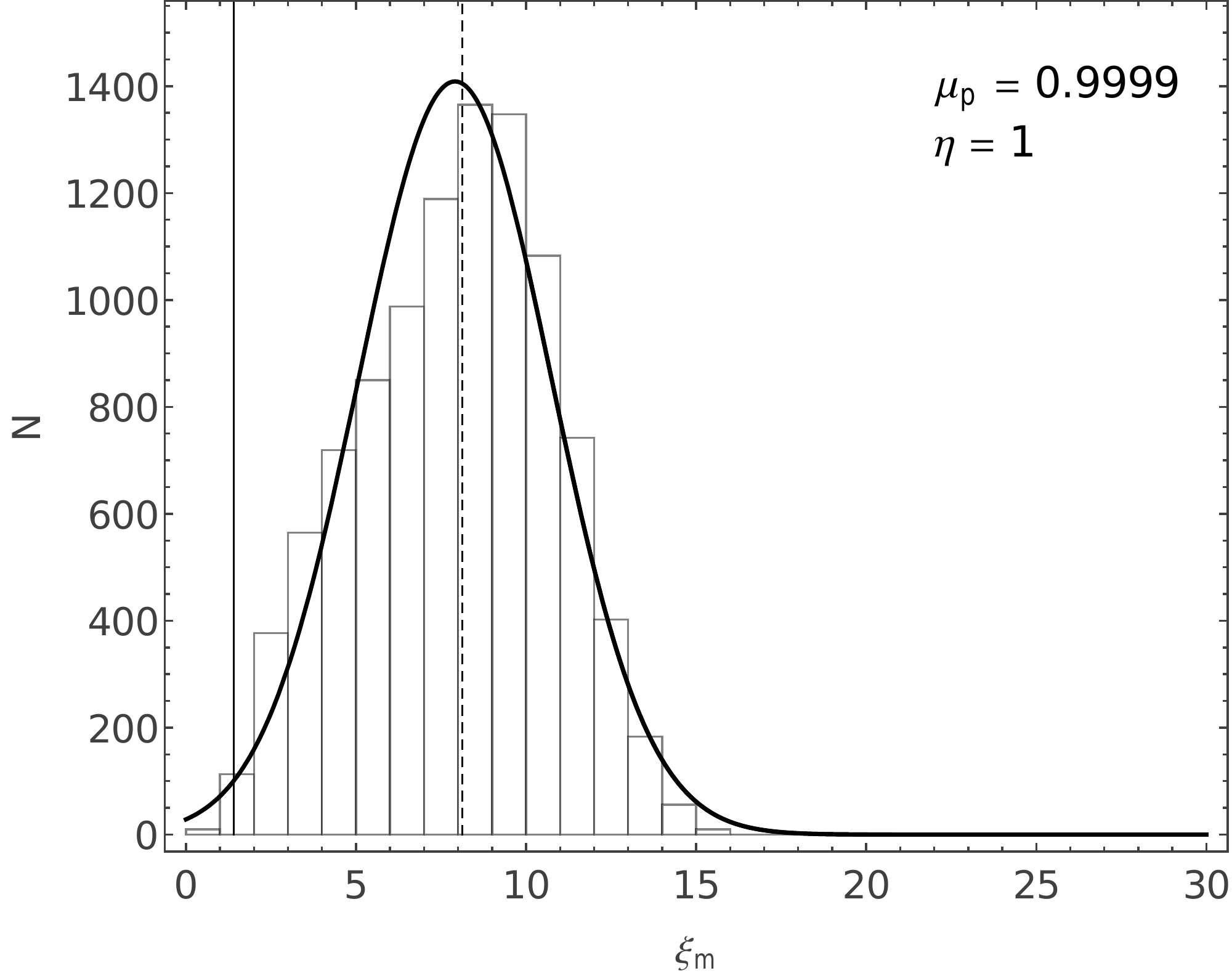} }}
\figcaption{Same as Figure 13, but for Experiment 8, for which $\cos\alpha_p = 0.9999$.
} 
\end{figure}

\begin{figure}
\figurenum{16}
{\centerline{\epsscale{0.80} \plotone{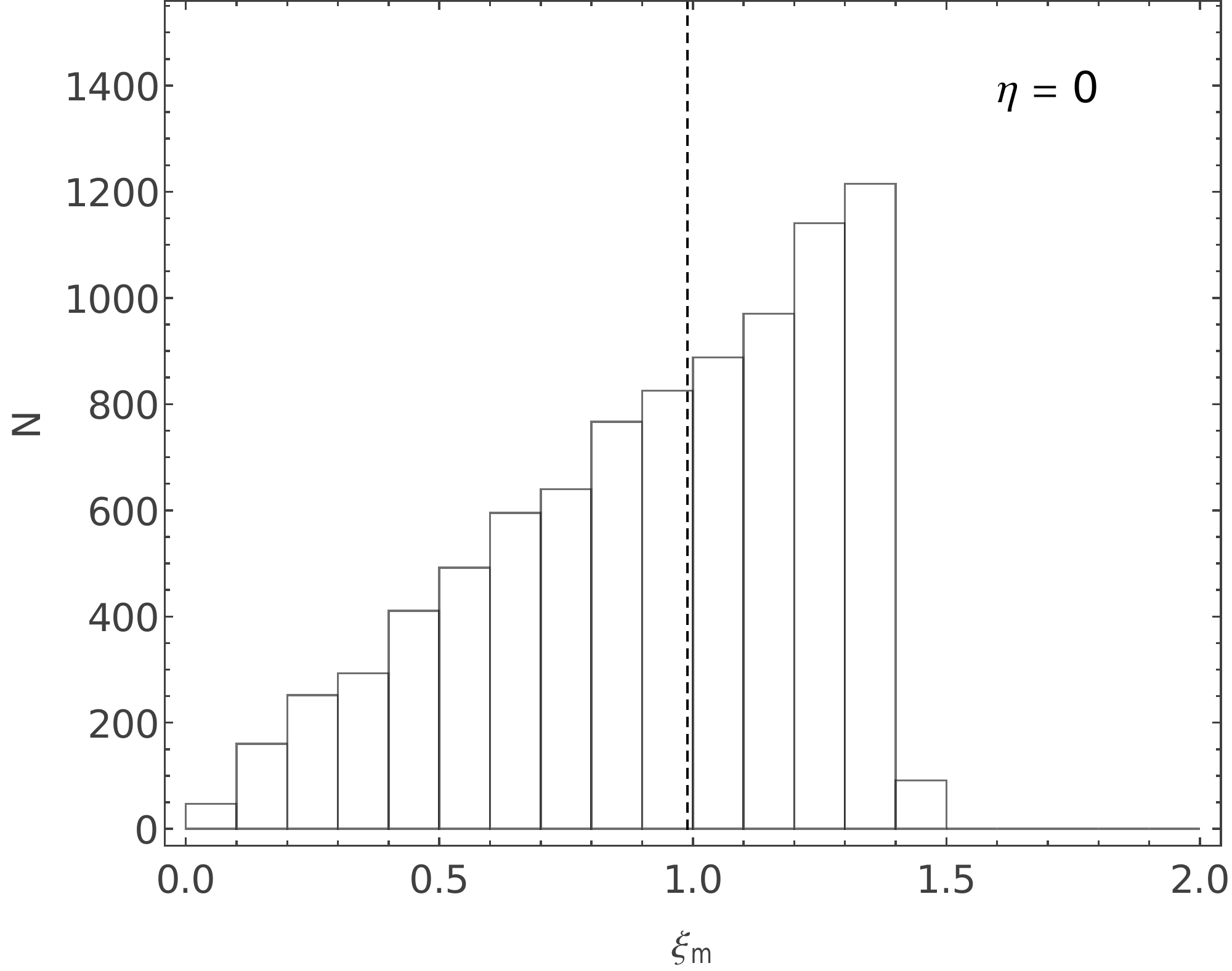} }}
\figcaption{Result of Experiment 9, as presented through  
the distribution of mirror radii for $10^3$ particles with Lorentz
factor $\gamma = 10^2$ and injected with pitch angle randomly selected
from a flat (uniform) distribution between $1\ge\mu_p\ge 0.9999$, and
from random locations at $z_i = 10^3 R$ from the part of $x-y$ plane
that funnels particles directly onto the inner surface.  The
background field $B_0$ is defined through the parameter $\varepsilon =
10^4$, and the turbulent magnetic field strength is set at $\eta = 0$
(no turbulent field).  The dashed line denotes the median value
$\bar\xi_m$ of distribution.  }
\end{figure}

\begin{figure}
\figurenum{17}
{\centerline{\epsscale{0.80} \plotone{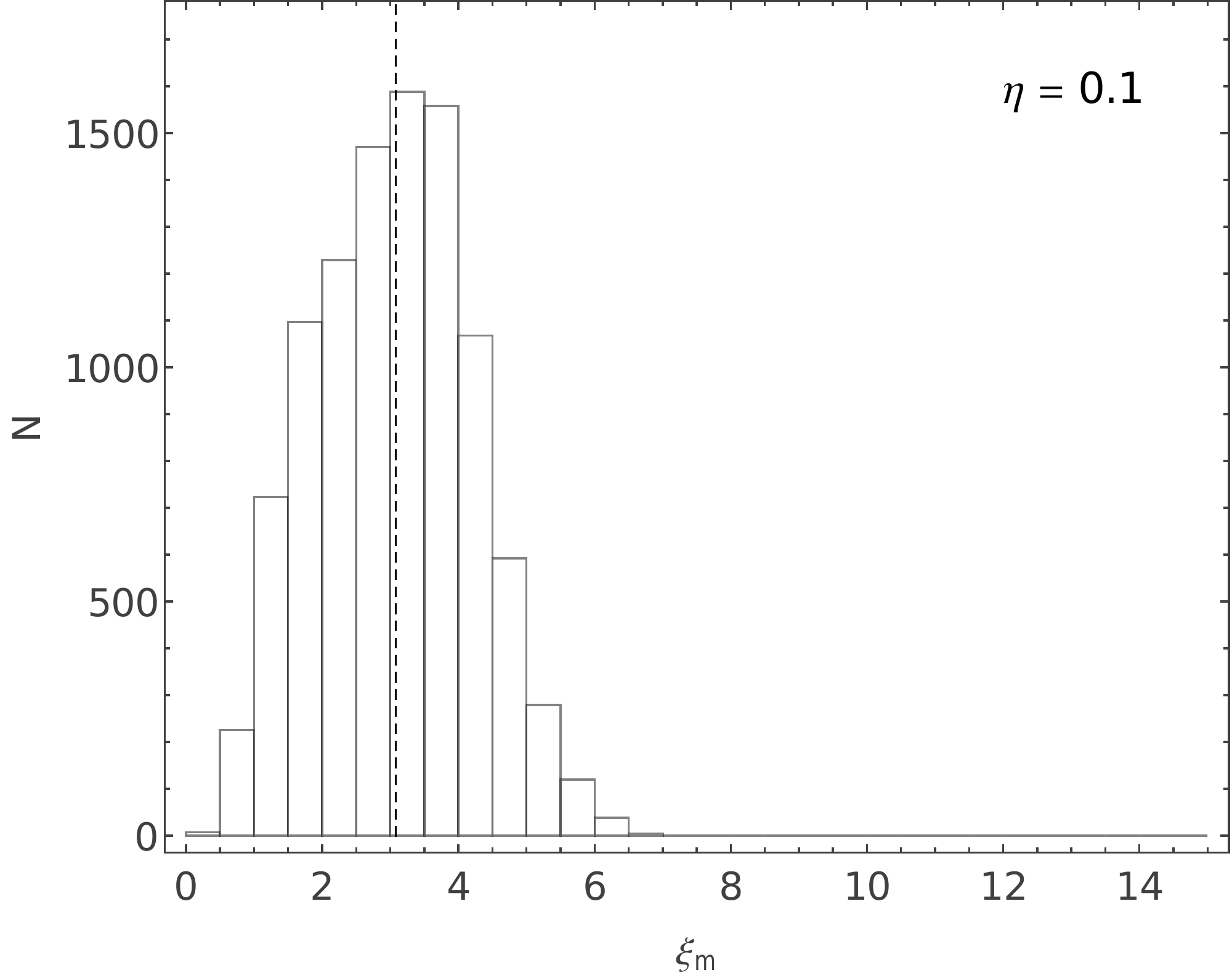} }}
\figcaption{Same as Figure 16, but for Experiment 10, for which $\eta = 0.1$.} 
\end{figure}

\begin{figure}
\figurenum{18}
{\centerline{\epsscale{0.80} \plotone{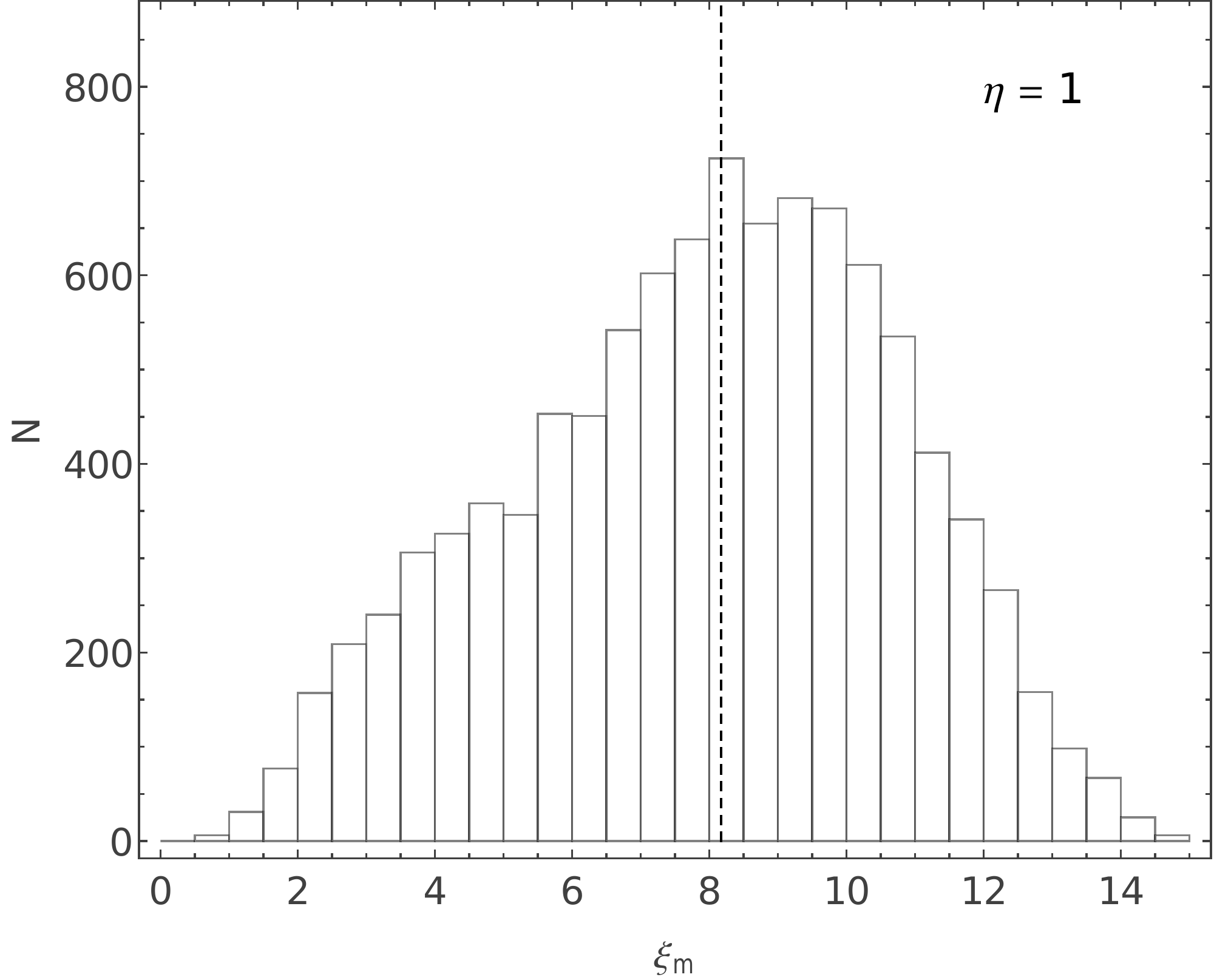} }}
\figcaption{Same as Figure 16, but for Experiment 11, for which  $\eta = 1$.
} 
\end{figure}

\begin{figure}
\figurenum{19}
{\centerline{\epsscale{0.80} \plotone{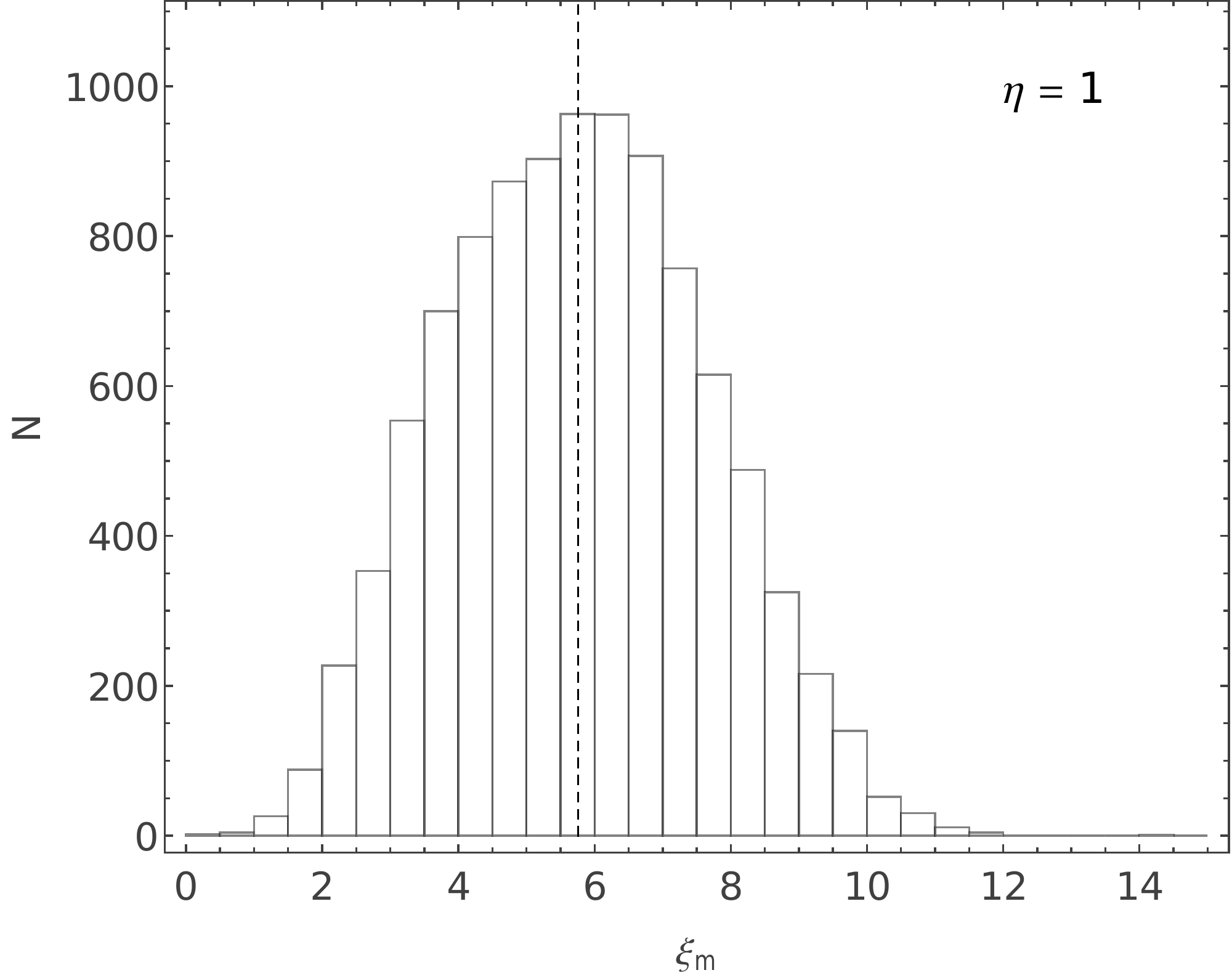} }}
\figcaption{Same as Figure 16, but for Experiment 12, for which $\eta = 1$ and $\gamma = 10$.
}  
\end{figure}

\begin{figure}
\figurenum{20}
{\centerline{\epsscale{0.80} \plotone{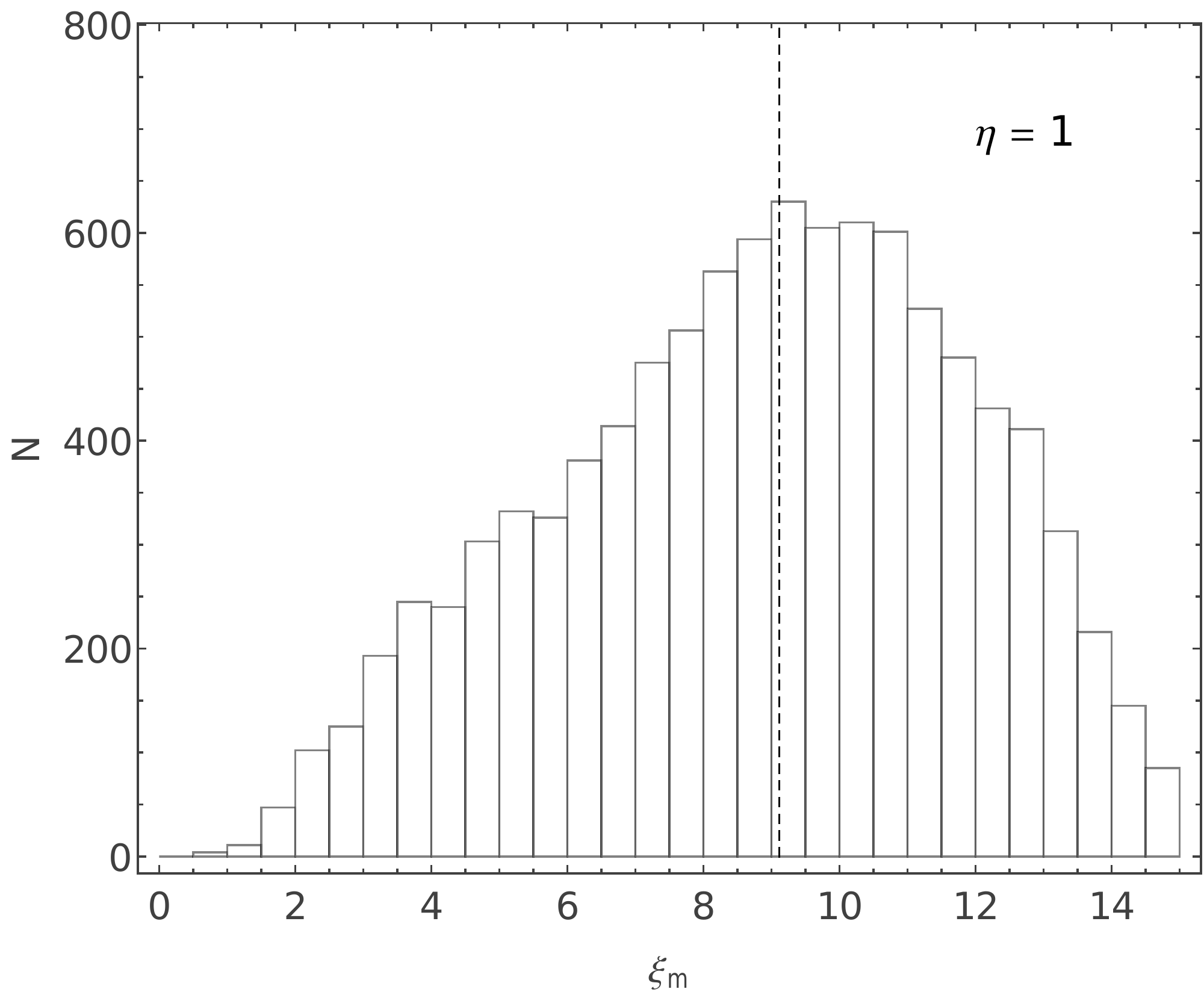} }}
\figcaption{Same as Figure 16, but for Experiment 13, for which $\eta = 1$ and  $\Gamma = 3/2$.
}  
\end{figure}

\section{Conclusion}

This paper considers the effects of turbulent fluctuations on the
propagation of cosmic rays impinging upon young star/disk systems. 
We focus on the case of magnetic fields with hour-glass-like
configurations and show how turbulence influences the magnetic
mirroring of incoming cosmic rays. The most important effects of
turbulence are to replace the mirroring point with a distribution of
values and to move the median mirror point outwards for sufficiently
large fluctuation amplitudes. More specifically, our results can be
summarized as follows:

We first construct a new coordinate system such that one coordinate
follows the magnetic field lines of the hour-glass configuration
(Section 2). The perpendicular coordinate is then used to construct
Alfv{\'e}nic field fluctuations, i.e., perturbations that are
perpendicular to the original magnetic field lines (Section 3). Using
the divergence operator of the new coordinate system, we can ensure
that the perturbations are divergence-free. 

Using this formulation of the problem, we have performed a large
number of numerical integrations for cosmic rays propagating along the
magnetic field lines, including the turbulent fluctuations. The
relevant parameter space is large: One must consider the field line in
question (labeled by its coordinate value $q$), the relative strength
$\varepsilon$ of the split-monopole and background field
contributions, the relative strength $\eta$ of the fluctuating field
components compared to the unperturbed field, as well as the initial
energy (given by the the Lorentz factors $\gamma$) and injection
inclination angle $\alpha_p$ of the cosmic rays. In addition, for each
choice of the variables $(q,\varepsilon,\eta,\gamma,\alpha_p)$, cosmic
rays will experience different realizations of the turbulent
fluctuations.  As a result, an ensemble of integrations must be
carried out for each set of starting conditions.  

The results of our numerical experiments (see Figures 8 -- 19) provide
us with the distribution of mirroring points for incoming cosmic rays
for given sets of initial conditions. Most notable, turbulence affects
the propagation of cosmic rays in these systems by replacing the
mirror point with a distribution of values.  If mirroring occurs at a
location in the field for which the magnitude of the turbulent
magnetic field component is small, $\delta B \ll 0.1 B_0$ (where $B_0$
is the magnitude of the underlying static field), the resulting
distribution is well described by a normal distribution with a median
value near the location of the mirror point found in the absence of
turbulence.  However, magnetic mirroring becomes enhanced once
particles enter a regime with larger fluctuation amplitudes with
$\delta B > 0.1 B_0$, even though the particles could penetrate
further into the turbulence-free field. The corresponding increase in
the median mirror point radius can be large, up to an order of
magnitude for the portion of parameter space considered herein (see
Table 1).  As a result, even a relatively modest amount of turbulence
($\eta \sim 0.1$) in young stellar objects can significantly reduce
the flux of cosmic rays reaching the disk.

A growing consensus in the field holds that disk accretion is produced
by an effective viscosity that is driven by turbulence, which in turn
is driven by MHD instabilities such as MRI \citep{balbushawley}. In
order for MRI to operate, and hence for disk accretion to take place,
the ionization fraction must be sufficiently high so that the gas is
well coupled to the field. The inner disk can be ionized by collisions
(where the number densities and temperatures are high), and the outer
disk can be ionized by standard values of the cosmic ray flux, but
intermediate regions may have dead zones where ionization is too low
\citep{gammie}. A reduced cosmic ray flux, such as that indicated
here, will thus act to decrease the fraction of the disk that is
active, i.e., sufficiently ionized for MRI to operate.  For
completeness we also note that T Tauri winds can also repel incoming
cosmic rays, in analogy to the Solar wind \citep{cleevestts}. As a
result, the cosmic ray flux could be too low for the disk to be MRI
active. In that case, the leading contribution to the ionization rate
is given by the decay of short-lived radioactive nuclei
\citep{umenakano,cleevesslr}. An important topic for additional work is
to ascertain how stellar winds and magnetic turbulence jointly
modulate the incoming cosmic ray flux, and how the result compares to
the contributions expected from radioactivity.

Another potential application of this work is to the magnetic braking
catastrophe, which can occur during the earlier protostellar stage of
evolution. In many circumstances, magnetic fields are so effective at
removing angular momentum from infalling protostellar envelopes that
circumstellar disks cannot form at all, or they are produced in highly
truncated configurations (for further detail see, e.g., \cite{li2013},
along with references therein). The failure to produce disks is a
theoretical problem, as observations indicate that circumstellar disks
are ubiquitous around young stellar objects. If protostellar systems
are sufficiently turbulent, however, magnetic field fluctuations can
increase the efficiancy of mirroring and thereby reduce the cosmic ray
flux in the inner region where disk formation takes place. With a
lower cosmic ray flux, and hence lower ionization levels, the gas will
be less well-coupled to the magnetic field, and magnetic braking can
be compromised. Although turbulence acts in the right direction to
alleviate the magnetic braking problem, further work must be carried
out to determine the size of the effect. In particular, turbulence
also acts to increase the rate of ambipolar diffusion
\citep{fat02,zweibel}, and the rate of magnetic reconnection
\citep{lazarian}, and both of these processes remove magnetic fields
from the inner collapse region and help facilitate disk formation. 
In addition, stochastic magnetic reconnection in a partially ionized
medium will also produce a magnetic cascade \citep{lazarian2004}. 
An important challenge for the future is to understand the interplay
between the reduction of ionization indicated here and the possible
increased rates of magnetic field diffusion and reconnection.

Finally, we note that this paper assumes that turbulence remains
robust in the inner regions of the hour-glass field configuration
where magnetic mirroring takes place. 
This assumption could be modified by ion-neutral damping, 
which acts to reduce the amplitude of magnetic turbulence 
when the frequency of magnetic waves is of the order of, 
or larger than, the ion-neutral collisional frequency.
If ion-neutral
damping is sufficiently effective, the turbulence levels in the inner
regions could be lower than assumed here. The degree of ion-neutral
damping depends on the magnetic field strength, the ionization levels,
the density of the background gas, and other parameters, all of which
have significant uncertainties, and all of which vary within the inner
regime of the young stellar object. As a result, the level of
ion-neutral damping, and the corresponding amplitudes for the
turbulence, are uncertain. Adding to the uncertainty, additional
sources of turbulence (e.g., from protostellar winds and outflows)
could also be operative. This paper parameterizes the degree of
turbulence through the parameter $\eta$ (see Table 1).  Another
challenge for the future is to develop a self-consistent model for the
turbulence for this inner region.

\acknowledgements

This work was supported by Xavier University through the Hauck Foundation. 
The authors thank the referee for useful comments that improved the
manuscript. 

$\,$

\end{document}